\pdfoutput=1
\documentclass[aps,prr,showpacs,twocolumn,superscriptaddress]{revtex4-1}
\usepackage{amsmath,latexsym,bm,psfrag,color,units}
\usepackage{graphicx}
\usepackage{epstopdf}
\usepackage{tabularx}
\usepackage{physics}
\usepackage{multirow}

\begin{document}

\title{$GW$ study of pressure-induced topological insulator transition in group IV-tellurides}

\author{ Pablo Aguado-Puente}
\affiliation{Atomistic Simulation Centre, School of Mathematics and Physics, Queen's University Belfast, Belfast BT7-1NN, Northern Ireland, United Kingdom}
\author{ Stephen Fahy }
\affiliation{Tyndall National Institute, Dyke Parade, Cork T12 R5CP, Ireland}
\affiliation{Department of Physics, University College Cork, College Road, Cork T12 K8AF, Ireland}
\author{Myrta Gr\"uning}
\affiliation{Atomistic Simulation Centre, School of Mathematics and Physics, Queen's University Belfast, Belfast BT7-1NN, Northern Ireland, United Kingdom}
\affiliation{European Theoretical Spectroscopy Facility}

\begin{abstract}
We calculate the electronic structure of the narrow gap semiconductors PbTe, SnTe and GeTe in the cubic phase using density functional theory (DFT) and the $G_0W_0$ method. Within DFT, we show that the band ordering obtained with a conventional semilocal exchange-correlation approximation is correct for SnTe and GeTe but wrong for PbTe. The correct band ordering at the high-symmetry point L is recovered adding $G_0W_0$ quasiparticle corrections. However, one-shot $G_0W_0$ produces artifacts in the band structure due to the wrong orbital character of the DFT single-particle states at the band edges close to L. We show that in order to correct these artifacts it is enough to consider the off-diagonal elements of the $G_0W_0$ self-energy corresponding to these states. We also investigate the pressure dependence of the band gap for these materials and the possibility of a transition from a trivial to a non-trivial topology of the band structure. For PbTe, we predict the band crossover and topological transition to occur at around 4.8 GPa. For GeTe, we estimate the topological transition to occur at 1.9 GPa in the constrained cubic phase, a pressure lower than the one of the structural phase transition from rombohedral to cubic. SnTe is a crystalline topological insulator at ambient pressure, and the transition into a trivial topology would take place under a volume expansion of approximately $10\%$.  
\end{abstract}

\date{\today}

\maketitle

\section{Introduction}
The triad of compounds, GeTe, SnTe and PbTe, constitute the basis for many materials with applications of great industrial interest, most notably thermoelectrics \cite{Rosi1968,Wood1988,Snyder2008} and phase-change materials \cite{Kolobov2004,Schneider2010}. Despite the historically widespread use of these materials, fundamental properties of their electronic structure are still receiving a lot of attention. In particular, SnTe was recently found to be the first realization \cite{Hsieh2012,Tanaka2012} of a new class of topological insulators in which the metallic surface states are protected by the crystal symmetry instead of the time-reversal symmetry \cite{Fu2011}. Interestingly, the isovalent counterparts, GeTe and PbTe, which share with SnTe their high temperature rocksalt atomic structure, do not display the band inversion that gives rise to the non-trivial topology of the electronic structure in SnTe. 

The band ordering in these materials is in fact governed by a delicate balance of interband interaction and spin-orbit coupling \cite{Ye2015}. They all display a small direct gap at L, with a second, very anisotropic hole pocket at $\Sigma$ very close below or above the top of the valence band at L. The characteristics of the band structure of these materials that stand out with respect to more conventional semiconductors, namely the very small direct gap at L (instead of $\Gamma$) and the possibility of a topological transition, can be linked to the unusual atomic character of the bands forming the top of the valence band (VB) and bottom of conduction band (CB) \cite{Hummer2007, Ye2015}. While in most semiconductors these states have a marked $sp$ character (bonding and antibonding in the case of elemental semiconductors such as silicon, or anion versus cation in the case of diatomic semiconductors such as GaAs), in the case of IV-VI semiconductors the top of the VB is mostly composed of states with anion $p$ orbital character, while the bottom of CB is formed by cation $p$ states \cite{Hummer2007}. The repulsion of these states at $\Gamma$ pushes the VB down, whereas at L repulsion from underlying cation $s$ states in the VB pushes the top of the VB up. This uplift is not counteracted by a repulsion from CB band states since the two band edges have opposite parity and therefore their interaction is forbidden by symmetry. 
It is precisely this lack of interaction between the band edges at L that potentially allows the band gap inversion and therefore a non-trivial topology of the band structure. 
The strong spin-orbit interaction in these materials further contributes to closing the gap by splitting the bands just below and above the top of the VB and bottom of CB respectively, pushing the band gap edges closer to the crossover point (or beyond in the case of SnTe).

The effect of pressure on the band structure of these materials is to enhance the $s-p$ repulsion of bands within the VB, giving rise to the usual negative sign of the pressure coefficient $\partial E_g/\partial P$ (where $E_g$ is the band gap) \cite{Svane2010a}. This means that, for the conventional narrow-gap semiconductors, PbTe and GeTe, pressure can be potentially used to close the gap and induce a topological transition. In this work, we explore this possibility by means of first-principles electronic-structure calculations and we provide an estimate of the critical pressures for the transition.

The proximity of these materials to the band crossover, and the dramatic changes in the band hybridization and topology that it entails, constitute a challenge for density functional theory (DFT) within the standard local or semilocal exchange-correlation approximations.
The band gap underestimation of local and semilocal functionals in DFT, which is caused by the overestimated delocalization of occupied states, can produce for these compounds a spurious inversion of the gap. Such is the case of PbTe, that local and semilocal DFT approximations predict to have the same band ordering at L as SnTe \cite{Hummer2007,Svane2010a,Robredo2019}, in contrast with experimental evidence.  
ARPES experiments in Pb$_x$Sn$_{1-x}$Te alloys demonstrate that PbTe and SnTe differ in their band ordering at L, \cite{Xu2012, Yan2014} in agreement with infrared spectroscopy measurements of the band gap variation with alloy composition \cite{Dimmock1966} and photoluminescence measurements of the sign of the deformation potentials \cite{Valeiko1991}.

It has been shown that the correct order of the bands in PbTe is recovered in DFT electronic structure calculation using hybrid \cite{Hummer2007} or meta-GGA \cite{Singh2013a,Ye2015} exchange and correlation approximations, or within a higher level of theory such as self-consistent quasiparticle $GW$ \cite{Svane2010a}. 
In general, hybrid DFT functionals have been successfully used for the study of group-IV chalcogenides \cite{Hummer2007,Murphy2018a}, but they have been shown to sometimes predict the wrong band ordering for materials close to topological transition and require confirmation from a higher order theory \cite{Svane2010a,Vidal2011}. 
Here, we use the $G_0W_0$ approach to correct the DFT electronic structure, including off-diagonal contributions to the self energy (Sec.~\ref{sc:method}),  and perform a comparative study of the effect of this correction on PbTe, SnTe and GeTe (Sec.~\ref{sec:electronic struct exp}). For PbTe, we show that $G_0W_0$ is sufficient to give the correct band ordering at L but off-diagonal contributions are needed to disentangle~\cite{VanSchilfgaarde2006} the incorrectly ordered DFT band structure around L.  
We discuss the possibility of transitions between a trivial and a non-trivial topology of the band structure with pressure, and calculate the value of the critical pressure (Sec.~\ref{sec:evolution with pressure}).

\section{Methods}

\subsection{Quasiparticle corrections}\label{sc:method}
The many-body quasi-particle energies and wave functions satisfy the equation
\begin{equation}
\begin{split}
      [\hat{T} + V_\mathrm{ext}(\bm{x}) + V_\mathrm{H}(\bm{x})]\psi_{\bm{k}n}(\bm{x}) + & \\
      \int{\Sigma(\bm{x},\bm{x}',E_{\bm{k}n})\psi_{\bm{k}n}(\bm{x}')d\bm{x}'} & = E^\mathrm{QP}_{\bm{k}n}\psi_{\bm{k}n}(\bm{x}),
\end{split}
\label{eq:QP}
\end{equation}
where the variable $\bm{x}$ contains both space and spin degrees of freedom. $\hat{T}$, $V_\mathrm{ext}(\bm{x})$ and $V_\mathrm{H}(\bm{x})$ are the kinetic energy operator, the external potential and the Hartree potential respectively. Within the GW approximation, the self-energy\cite{Aryasetiawan1998} is  
\begin{equation}
\begin{split}
    & \Sigma(\bm{x},\bm{x}',\omega) = \\
    & {i \over 2 \pi} \lim_{\eta\to 0} \int{ e^{i\eta\omega'}G(\bm{x},\bm{x}',\omega-\omega')W(\bm{r},\bm{r}',\omega')d\omega'}.
\end{split}
\label{eq:GW}
\end{equation}
In the standard $G_0W_0$ approximation on top of DFT, the Green's function $G$ and the screening $W$ in Eq.~\ref{eq:GW} are calculated using the DFT energies $\{ \varepsilon^\mathrm{DFT}_{\bm{k}n}\}$ and wavefunctions $\{\phi^\mathrm{DFT}_{\bm{k}n}\}$. Eq.~\ref{eq:QP} is solved to first-order in the perturbation $\Sigma - V_{xc}$, where $V_{xc}$ is the exchange-correlation potential from DFT, giving quasiparticle energies 
\begin{equation}
    E^\mathrm{QP}_{\bm{k}n} = \varepsilon^\mathrm{DFT}_{\bm{k}n} + \mel{\phi^\mathrm{DFT}_{\bm{k}n}}{\Sigma(E^\mathrm{QP}_{\bm{k}n})  - V_{xc}}{\phi^\mathrm{DFT}_{\bm{k}n}}.
\label{eq:self-energy}
\end{equation}
At the first-order, the quasiparticle wavefunctions (Eq.~\ref{eq:QP}) are approximated by the DFT ones, which turns to be a satisfactory approximation for many $sp$ systems. However, for narrow band-gap semiconductors, DFT can give the wrong band-ordering for states close to the Fermi energy, as it is the case for PbTe at L or e.g. of Ge at $\Gamma$~\cite{VanSchilfgaarde2006}. As a consequence, the corresponding DFT wavefunctions can have the wrong orbital character: in this case a diagonal-only $G_0W_0$ correction as in Eq.~\ref{eq:self-energy} is not sufficient, an issue referred to as the ``band disentanglement problem'' in Ref.~\onlinecite{VanSchilfgaarde2006}.
Using a better reference than (semi)local DFT, e.g. hybrid DFT, or resorting to self-consistent quasi-particle $GW$, in which both the quasi-particle energies and wave-functions are updated self-consistently, both solve this issue. Here, we use instead a relatively inexpensive extension of the diagonal-only $G_0W_0$ approximation for which we compute the $G_0W_0$ self-energy corrections over a small sub-space of relevant states, including the off-diagonal elements:
%
%
%
\begin{equation}
\begin{split}
     H^{GW}_{nm}(\bm{k};E^\mathrm{QP}_{\bm{k}n}) = & \; \varepsilon^\mathrm{DFT}_{\bm{k}n}\delta_{nm}
      + \Delta \Sigma_{nm},
\end{split}
\label{eq:offdiag}
\end{equation}
where 
\begin{equation}
\Delta \Sigma_{nm} = \mel{\phi^\mathrm{DFT}_{\bm{k}n}}{\Sigma(E^\mathrm{QP}_{\bm{k}n})  - V_{xc}}{\phi^\mathrm{DFT}_{\bm{k}m}}    
\end{equation}
The corrections to the DFT eigenvalues are found by diagonalization after linearization and hermitization of the matrix (see Appendix~\ref{sc:app}). Similar approaches were applied successfully in systems such as topological insulators \cite{Aguilera2013,Forster2016}---for which small corrections greatly affect the mixing of the bands near the band-crossing points---and materials with a strong $p-d$ hybridization.~\cite{VanSchilfgaarde2006}

\subsection{Computational details}
The starting point for our $GW$ calculations are density functional theory (DFT) simulations carried out with the Quantum-{\sc espresso} suite \cite{Giannozzi2009,Giannozzi2017}. DFT calculation were performed within the generalized gradient approximation (GGA) using the Perdew-Burke-Ernzerhof (PBE) \cite{Perdew1996} parametrization of the exchange and correlation functional. We have used fully nonlocal two-projector norm-conserving pseudopotentials from the {\sc PseudoDojo} data base \cite{VanSetten2018} generated with the ONCVPSP code \cite{Hamann2013}. All pseudopotentials included a full semicore shell. Wave functions in the DFT calculation were expanded in a basis of plane waves with cutoffs of 110, 100 and 100 Ry for GeTe, SnTe and PbTe respectively. Brillouin zone sampling was carried out using a Monkhorst-Pack mesh \cite{Monkhorst1976} of $12\times12\times12$ for the self-consistent calculations. All calculations include spin-orbit interaction.

The $GW$ corrections were computed using the Yambo code \cite{Marini2009,Sangalli2019} ---modified to calculate the off-diagonal elements of the self-energy. We used a $16\times16\times16$ grid on the Brillouin zone. A cutoff of 30 Ry for the $\bm{G}$-vectors was used for the computation of the exchange-self energy. The correlation part of the self-energy was computed summing over 100 bands (46 occupied), and the screening was calculated within the plasmon-pole approximation using 120 bands. We use the standard diagonal-only $G_0W_0$ approximation except when otherwise specified. 

\section{Results}
\subsection{Electronic structure at the experimental lattice parameter}
\label{sec:electronic struct exp}

\begin{figure*}[]
    \begin{center}
        \includegraphics[width=0.66\columnwidth]{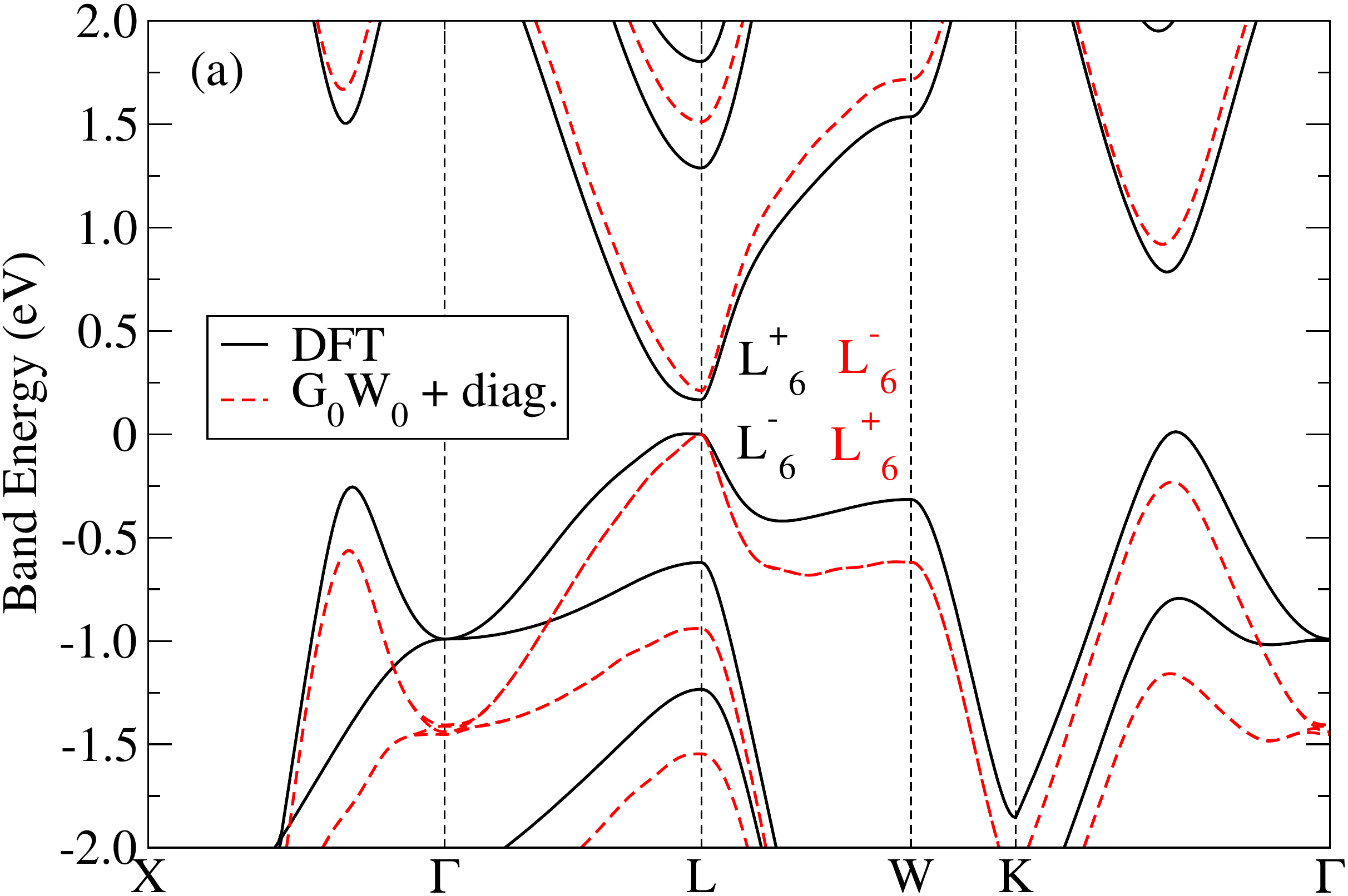}
        \includegraphics[width=0.66\columnwidth]{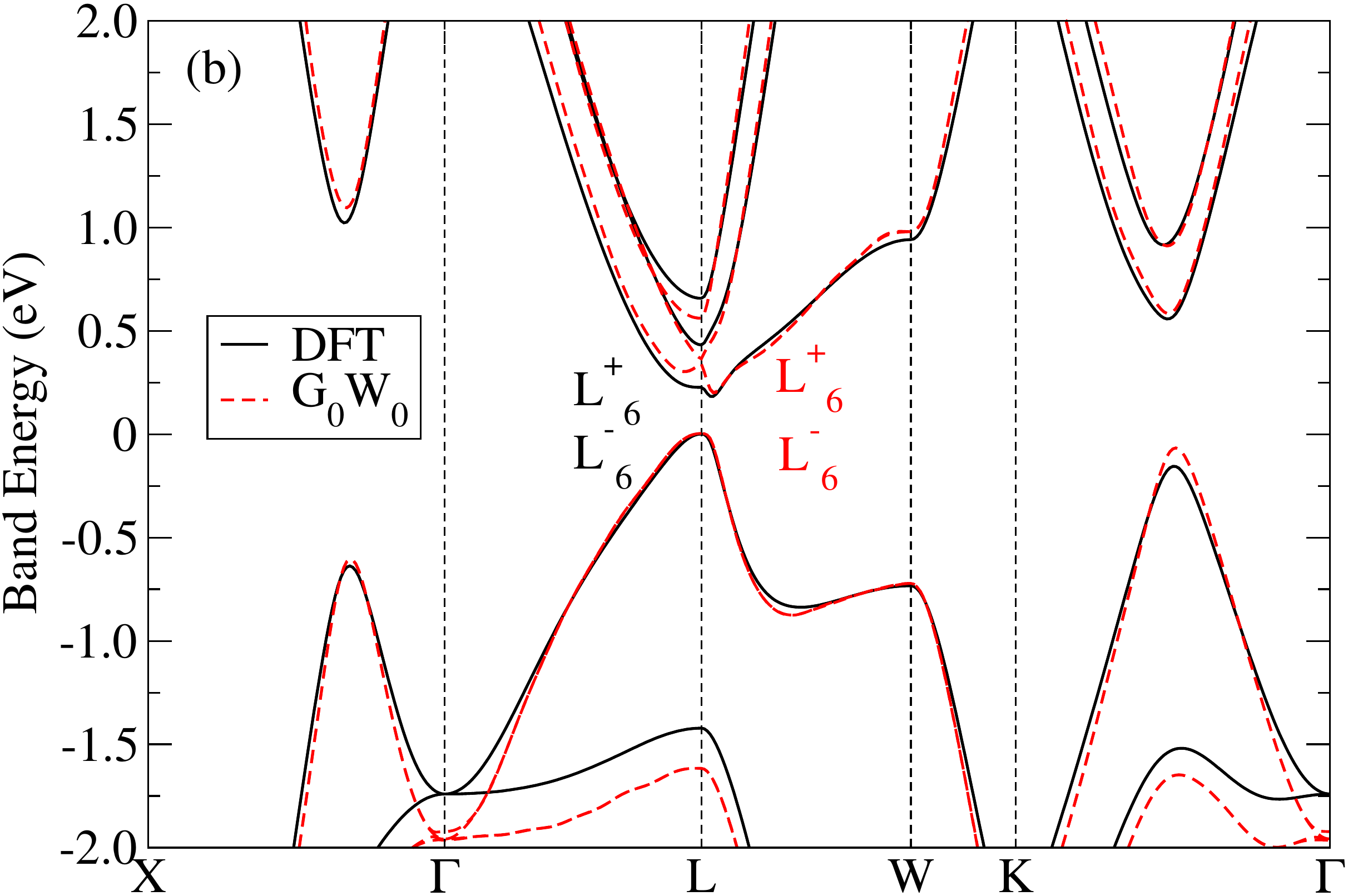}
        \includegraphics[width=0.66\columnwidth]{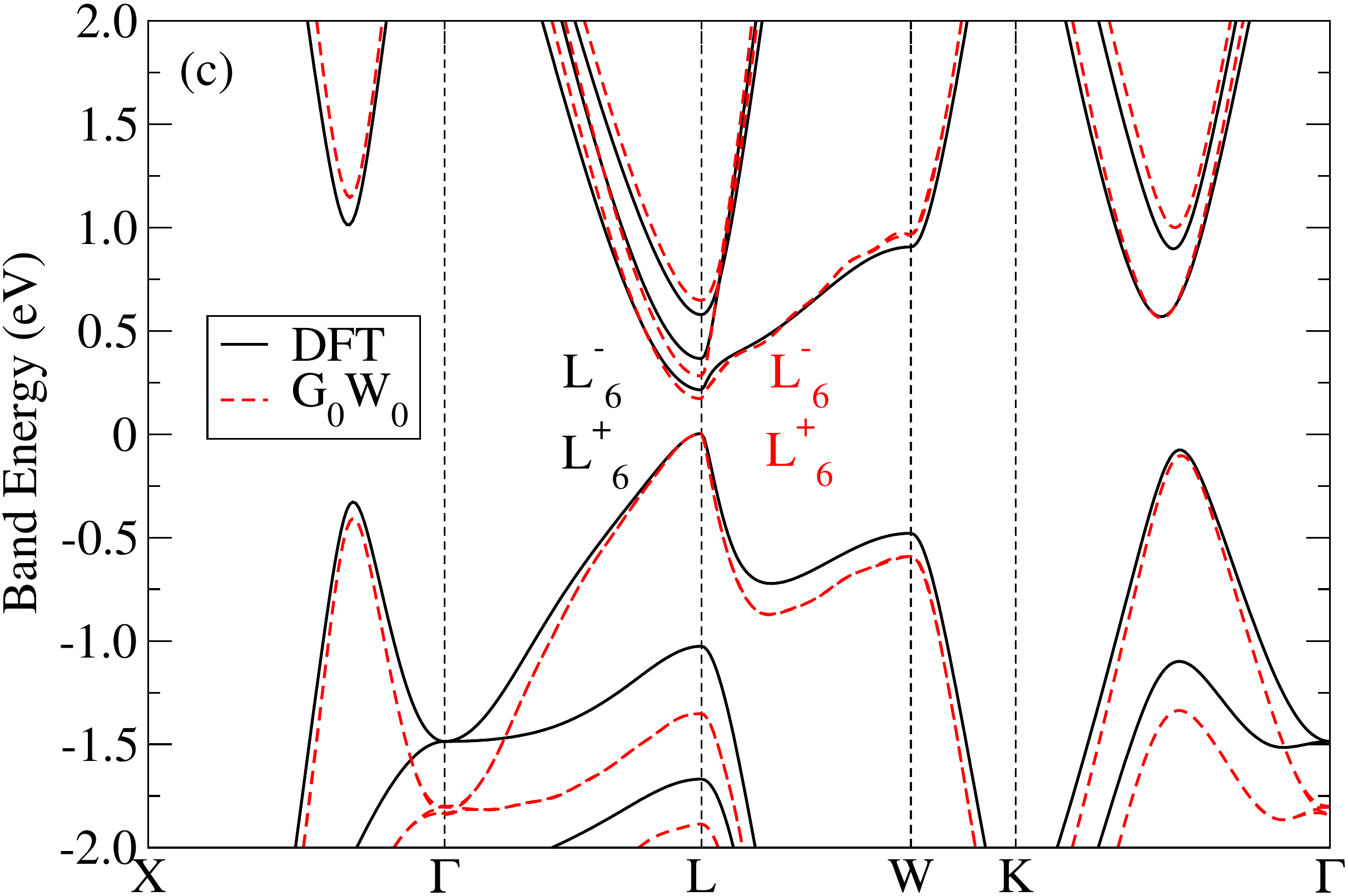}
        
        \caption{Band structure near the direct band gap at L for (a) PbTe, (b) SnTe and (c) GeTe in the rocksalt structure, as obtained with GGA DFT (solid black lines) and $G_0W_0$ (dashed red lines). These band structures were obtained at the experimental lattice parameters for the cubic phase: 6.462, 6.327 and 6.01 {\AA} for PbTe \cite{MadelungHandbook}, SnTe \cite{MadelungHandbook} and GeTe \cite{Wiedemeier1977,Chattopadhyay1987}, respectively }
        \label{fig:bandStructure}
    \end{center}
\end{figure*}

We show in Fig. \ref{fig:bandStructure} a detail of the band structure near the direct band gap at L for PbTe, SnTe and GeTe obtained at the experimental lattice parameters for the cubic phase: 6.462 \cite{MadelungHandbook}, 6.327 \cite{MadelungHandbook} and 6.01 \AA \cite{Wiedemeier1977,Chattopadhyay1987}, respectively. 
It should be noted that while for PbTe the NaCl structure constitutes the ground state of the system in ambient conditions, both SnTe and GeTe present a distorted rhombohedral R3m structure at low temperature. SnTe may crystallize in the cubic phase in Te-rich conditions \cite{Brebrick1971}. The rocksalt structure in GeTe is only stable above 400$^\circ$ C \cite{Shu1987}.
All three material have very similar band structures in the cubic phase. 

For PbTe our band structure calculated with the PBE functional displays the spurious inverted band gap (L$^-_6-$L$^+_6 <0$), in agreement with previous reports with local and semilocal DFT functionals \cite{Hummer2007, Svane2010a}.

\begin{figure}[]
    \begin{center}
        \includegraphics[width=0.9\columnwidth]{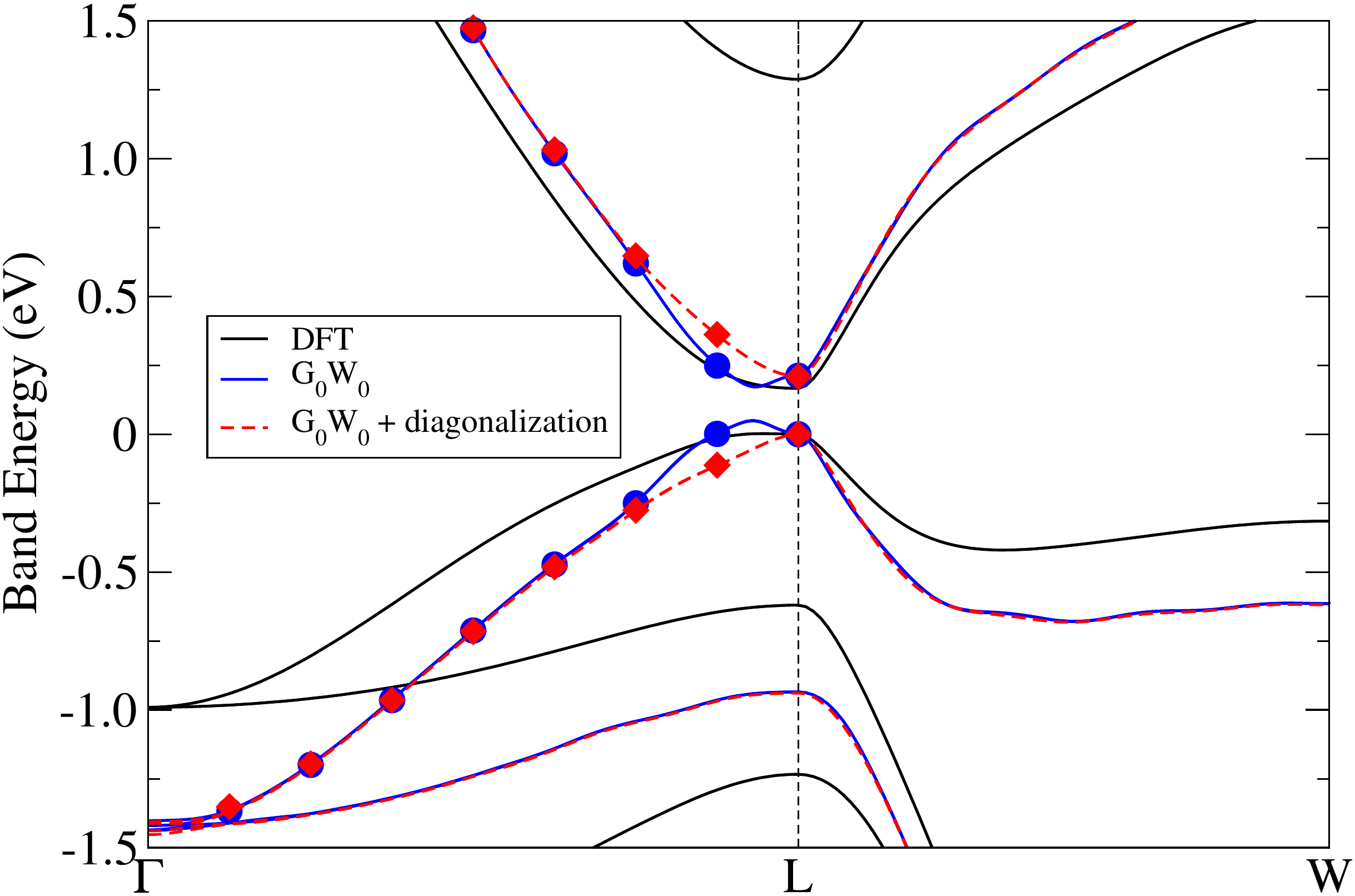}
        \caption{Band structure near the direct band gap at L for PbTe. In black we plot results obtained with the PBE functional of DFT, in blue those obtained with conventional diagonal $G_0W_0$ and in red those calculated after diagonalizing the $G_0W_0$ Hamiltonian. For the $GW$ bands, symbols correspond to $k$-points for which the correction was computed, while lines were obtained by Wannier interpolation.}
        \label{fig:offdiag}
    \end{center}
\end{figure}

We use $G_0W_0$ to obtain the corrected electronic band structures. We have found that the conventional diagonal-only $G_0W_0$ is sufficient to correct the order of the bands at L, where any spurious mixing of the L$^-_6$ and L$^+_6$ states is forbidden by symmetry and therefore, other than the wrong ordering of the eigenvalues, the DFT wavefunctions
are a good representation of the real wave functions of the states. 
On the other hand, we have observed that the diagonal-only $G_0W_0$ band structure presents artifacts for $k$-points close but different from L when the $G_0W_0$ correction involves a change in sign of the $L_6^- - L_6^+$ gap. For the band structure of PbTe at the experimental cell volume, these are shown in Fig. \ref{fig:offdiag}. In fact, the spurious inversion of the gap produced by DFT causes a mixing of the orbital character of the bands that is virtually non-existent in the case of a non-inverted gap \cite{Hummer2007}. Since it relies on the DFT wavefunctions (Sec.~\ref{sc:method}), conventional diagonal-only $G_0W_0$ is then unable to disentangle the DFT band structure and this results in the artifacts at the band edges close to L.

Figure \ref{fig:offdiag} shows how these artifacts are corrected after calculating the off-diagonal matrix elements from Eq.~\ref{eq:offdiag} and diagonalizing over a subspace comprising just the four bands forming the edges of the conduction and valence bands (taking into account spin degeneracy). Diagonalizing a larger subspace, comprising up to 12 bands has no further effect on the correction. 

We turn now our attention to SnTe, a system in which, according to ARPES experiments \cite{Tanaka2012,Xu2017} and {\it ab initio} simulations \cite{Hsieh2012,Druppel2014,Lee2016} the inversion of the band gap does occur. Our DFT band structure, shown in black in Fig. \ref{fig:bandStructure}(b), displays an inverted band gap (L$^-_6-$L$^+_6 <0$), in agreement with previous reports. In this case the $G_0W_0$ correction, shown in red in Fig. \ref{fig:bandStructure}(b) increases the size of the \emph{negative} gap bringing it closer to the low temperature experimental value.  
In the case of SnTe, DFT gives the correct band ordering,
changes in the bands hybridization are not expected to be as significant as in PbTe, and we find indeed that the correction from the off-diagonal elements is negligible in this case (the absolute value of the off-diagonal elements of the self energy matrix are smaller than 1\% of the values of the diagonal ones).

For the sake of completeness we have also performed the same analysis for cubic GeTe. For GeTe we have not been able to find any experimental evidence for the sign of the $L_6^- - L_6^+$ band gap. Whereas the pioneering electronic structure calculations of Cohen and coworkers using empirical pseudopotentials predicted for GeTe the same band ordering as for SnTe~\cite{Tung1969}, contemporary DFT simulations using meta-GGA functionals suggest a non-inverted scenario \cite{Ye2015, Singh2013a}. Our calculations, both with GGA DFT and including the $G_0W_0$ correction, agree with the latter result, giving a trivial topology of the band structure, similar to that of PbTe. In this case too the off-diagonal correction is negligible, as expected, since the $GW$ correction does not reverse the order of the bands at L.

In Table \ref{tab:masses} we compile the effective masses for the three materials computed using the band structures obtained with both DFT and $G_0W_0$. 
For PbTe, the only material for which experimental data could be found, the overestimation of the effective masses by DFT due to the spurious hybridization of valence and conduction band near L is fixed by $G_0W_0$, producing values in much better agreement with the experiment. 
For SnTe we only report values for the valence band, since the proximity of the band gap inversion crossover makes the dispersion of the conduction band very non-parabolic near the L point, and therefore a value of the effective mass cannot be determined. $G_0W_0$ tends to slightly increase the effective masses of holes.
In the case of GeTe, the corrections of $G_0W_0$ over the values obtained with DFT are much smaller than for SnTe and specially PbTe.
We were not able to find experimental reports of the effective masses for neither SnTe nor GeTe.

\begin{table}[]
   \begin{tabularx}{\columnwidth}{ll|XXX}
    \hline \hline
            & & DFT & $G_0W_0$ & exp. \\
    \hline
        \multirow[t]{4}{*}{PbTe} &  $m_\parallel^\mathrm{h}$  & -0.69  & -0.37  & -0.31 \\
                                  &  $m_\perp^\mathrm{h}$     & -0.047 & -0.034 & -0.022 \\
                                  &  $m_\parallel^\mathrm{e}$ &  0.52  &  0.24   &  0.24 \\
                                  &  $m_\perp^\mathrm{e}$     &  0.035 &  0.034 &  0.024 \\
        \hline \\
        \multirow[t]{4}{*}{SnTe}  &  $m_\parallel^\mathrm{h}$ & -0.34  & -0.46 & -- \\
                                  &  $m_\perp^\mathrm{h}$     & -0.075 & -0.12 & -- \\
                                  &  $m_\parallel^\mathrm{e}$ &  n/a & n/a & -- \\
                                  &  $m_\perp^\mathrm{e}$     &  n/a & n/a & -- \\
        \hline \\
        \multirow[t]{4}{*}{GeTe}  &  $m_\parallel^\mathrm{h}$ & -0.39  & -0.39  & -- \\
                                  &  $m_\perp^\mathrm{h}$     & -0.022 & -0.014 & -- \\
                                  &  $m_\parallel^\mathrm{e}$ &  0.49  &  0.45  & -- \\
                                  &  $m_\perp^\mathrm{e}$     &  0.024 &  0.019 & -- \\
    \hline \hline
    \end{tabularx}
    \caption{Effective masses for cubic PbTe, SnTe and GeTe at the experimental volumes as obtained with DFT and $G_0W_0$ (including off-diagonal corrections). The reason for not reporting the values for electrons in the case of SnTe is explained in the text. We compare with the experimental values for PbTe \cite{Dalven1969}, the only material for which the corresponding data could be found in the literature.  }
    \label{tab:masses}
\end{table}

\subsection{Evolution of gap with volume and topological crossover}
\label{sec:evolution with pressure}
\begin{figure*} []
    \begin{center}
        \includegraphics[width=0.66\columnwidth]{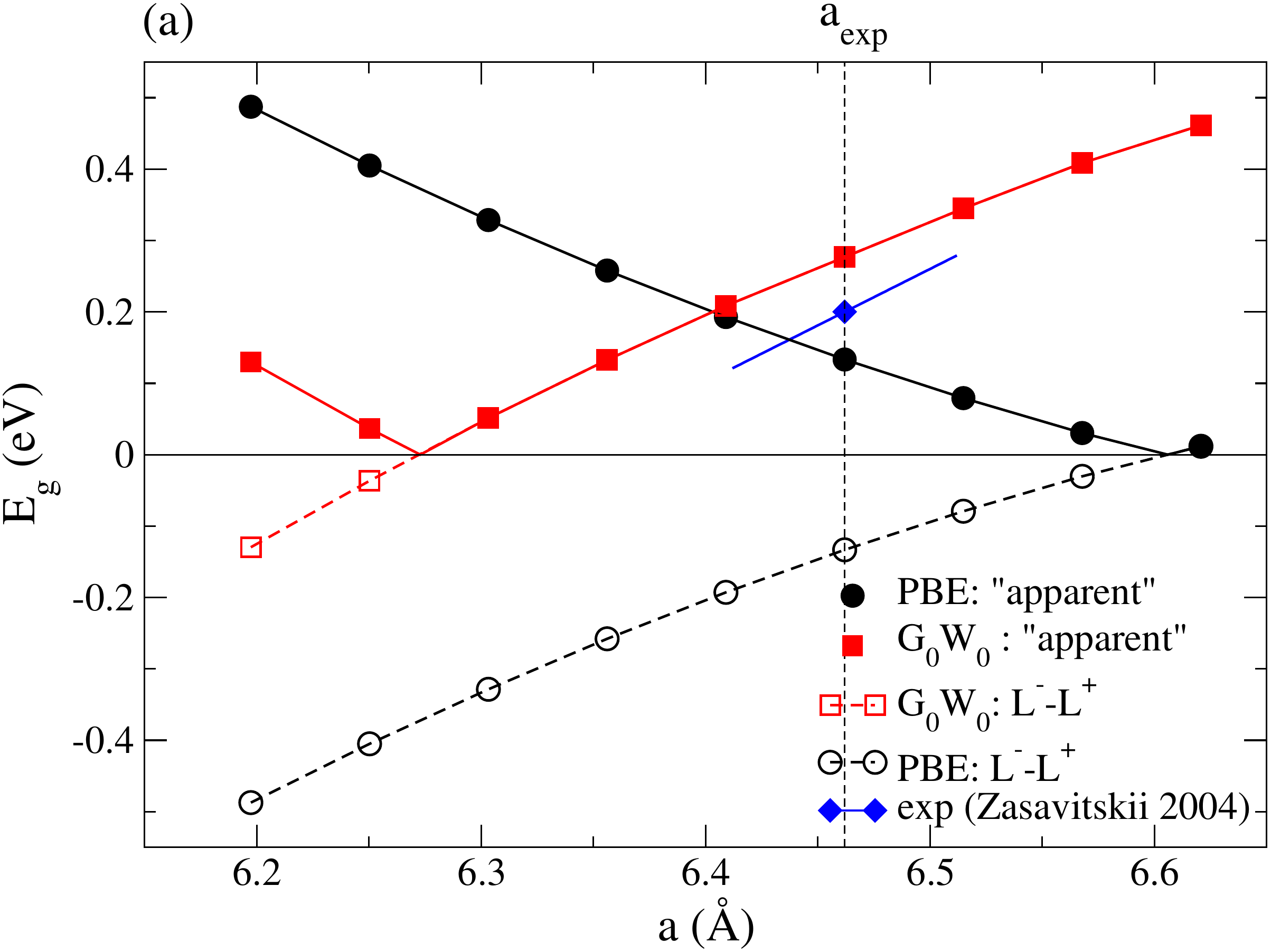}
        \includegraphics[width=0.66\columnwidth]{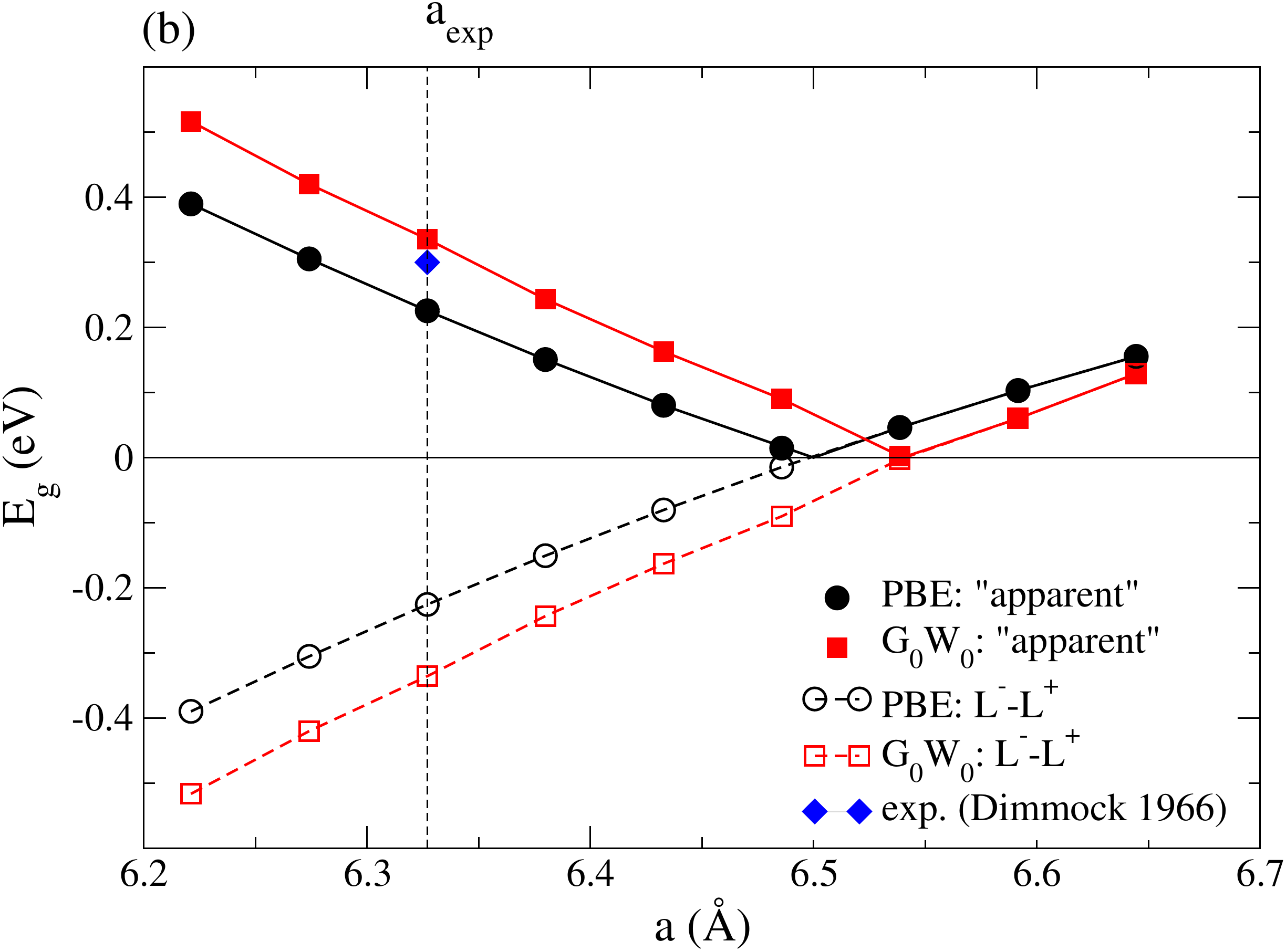}
        \includegraphics[width=0.66\columnwidth]{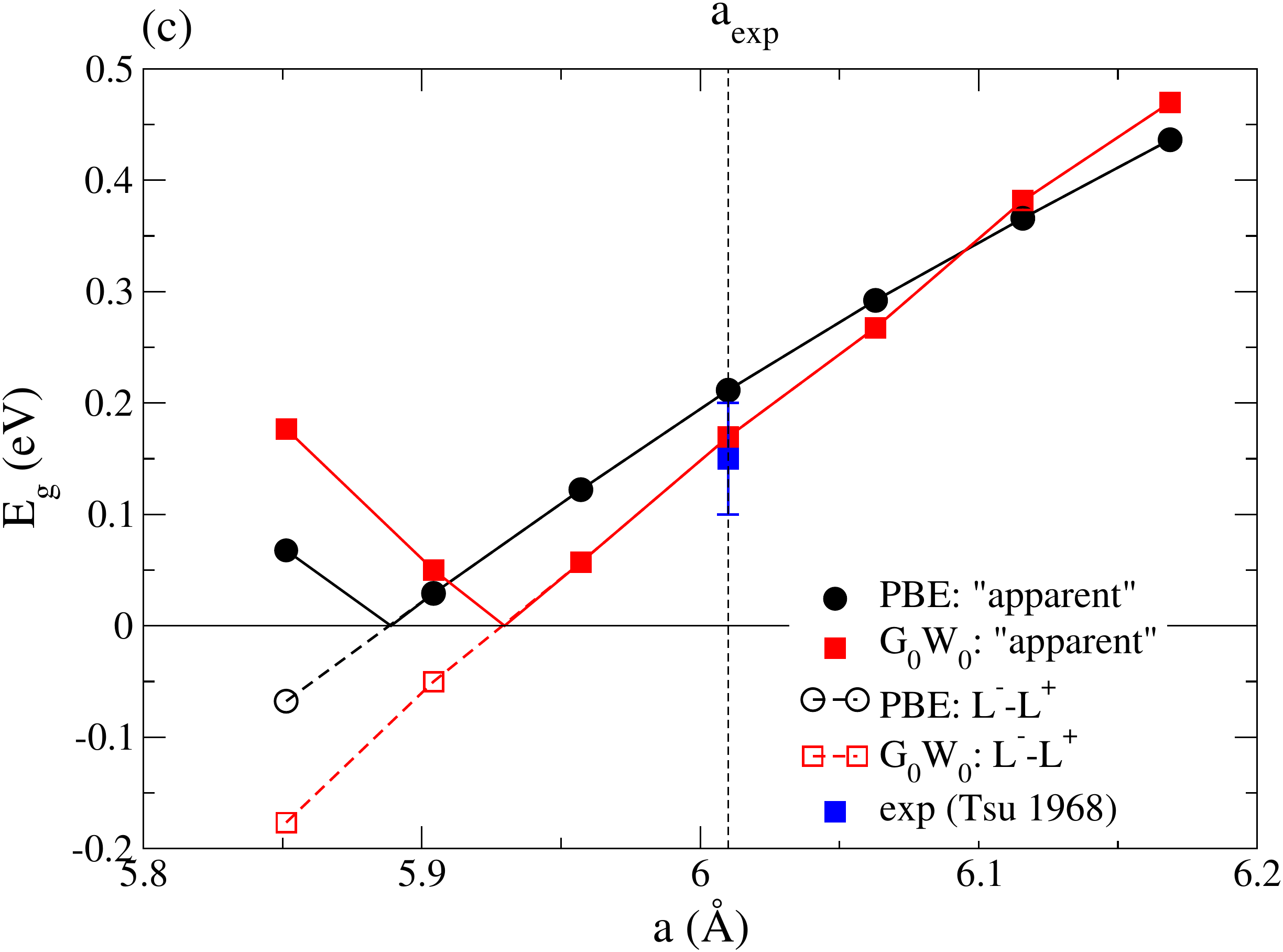}
        
        \caption{Band gap as a function of lattice parameter for (a) PbTe, (b) SnTe and (c) GeTe in the rocksalt structure. Solid lines represent the apparent band gap (lowest unoccupied state minus highest occupied state) at L as obtained with GGA DFT (black) and $G_0W_0$ (red). Dashed lines show the band gap calculated as the energy difference between L$_6^--$L$_6^+$, again as obtained with GGA (black) and $G_0W_0$ (red). Experimental values of the band gap are shown in blue, when available we also display the experimental error bar or the derivative with volume.}
        \label{fig:gapVsAlat}
    \end{center}
\end{figure*}

It has been previously suggested --- although, to our knowledge, never demonstrated ---that, because of (i) the lack of interaction between the $L_6^-$ and $L_6^+$ states at L and (ii) the unusual negative sign of the pressure coefficients, $\partial E_g/\partial P$, it should be possible to drive the electronic structure of some of these materials through a topological transition with pressure~\cite{Robredo2019, Svane2010a}.

We plot in Fig.~\ref{fig:gapVsAlat} the evolution of the direct band gap at L with the volume for PbTe, SnTe and GeTe, calculated both at the level of DFT and $G_0W_0$ approximation. 
As already discussed, SnTe is on the non-trivial topology side of the crossover at ambient pressure, therefore by applying positive pressure (reducing the volume of the unit cell) one increases the magnitude of the already negative gap, as shown by the dashed lines in Fig. \ref{fig:gapVsAlat}(b). Instead, according to our $G_0W_0$ calculations, inducing the transition to a trivial insulator would require a volume expansion of the material by 10\%. One of the effects of alloying SnTe with PbTe is in fact the volume expansion of the material, which contributes to the observed band crossover in this alloy \cite{Xu2012}.
Thermal expansion would also bring this material closer to the topological transition. Considering the critical volume obtained from Fig.~\ref{fig:gapVsAlat}(b) and the thermal expansion coefficient of SnTe, $\alpha_V \sim 6\cdot 10^{-5}$ K$^{-1}$ ~\cite{MadelungHandbook}, would yield a temperature for the band gap closing of $\sim 1700$K, however this gross estimate is neglecting the band gap renormalization from electron-phonon interaction which may significantly reduce this temperature \cite{Allen1976,Allen1981,Cardona1983}.

\begin{table}[]
    \begin{tabularx}{\columnwidth}{X|XXX}
    \hline \hline
            & PbTe & SnTe & GeTe \\
    \hline
        DFT        & -2.5 & -2.7 & 3.0 \\
        $G_0W_0$   &  4.8 & -3.2 & 1.9 \\
    \hline \hline
    \end{tabularx}
    \caption{Crystalline topological insulator transition pressures in GPa. Critical pressures have been obtained from the critical lattice constants in Fig. \ref{fig:gapVsAlat} using the Murnaghan equation of state. }
    \label{tab:transitionPressures}
\end{table}

More interesting is the case of PbTe. Fig.~\ref{fig:gapVsAlat}(a) shows the dependence of its gap with volume. According to the DFT calculations this material already possesses a topologically non-trivial band structure, and thus, like in SnTe it would have only been possible to induce the topological transition by the application of negative pressure, expanding its volume by 7\%. Instead, the $G_0W_0$ corrections to the band structure puts this material on the other side of the transition, as a trivial insulator, in agreement with the experiments. Taking into account this correction the transition would occur for a positive pressure, at a volume compression of 9\%. 
From this volume compression an estimate for the crossover pressure can obtained using the Murnaghan equation of state \cite{Murnaghan1944, Tyuterev2006} fitted to a series of calculations of total energy values as a function of volume. The parameters of these fits can be found in Table \ref{tab:Murnaghan} for all three materials. For PbTe this yields an estimate for the transition pressure from trivial insulator to a topological one of around 4.8 GPa.
It should be noted that due to the inverted gap obtained by the DFT calculations, within this approach one would obtain a topological transition at a negative pressure of -2.5 GPa. 

\begin{figure}[]
    \begin{center}
        \includegraphics[width=0.9\columnwidth]{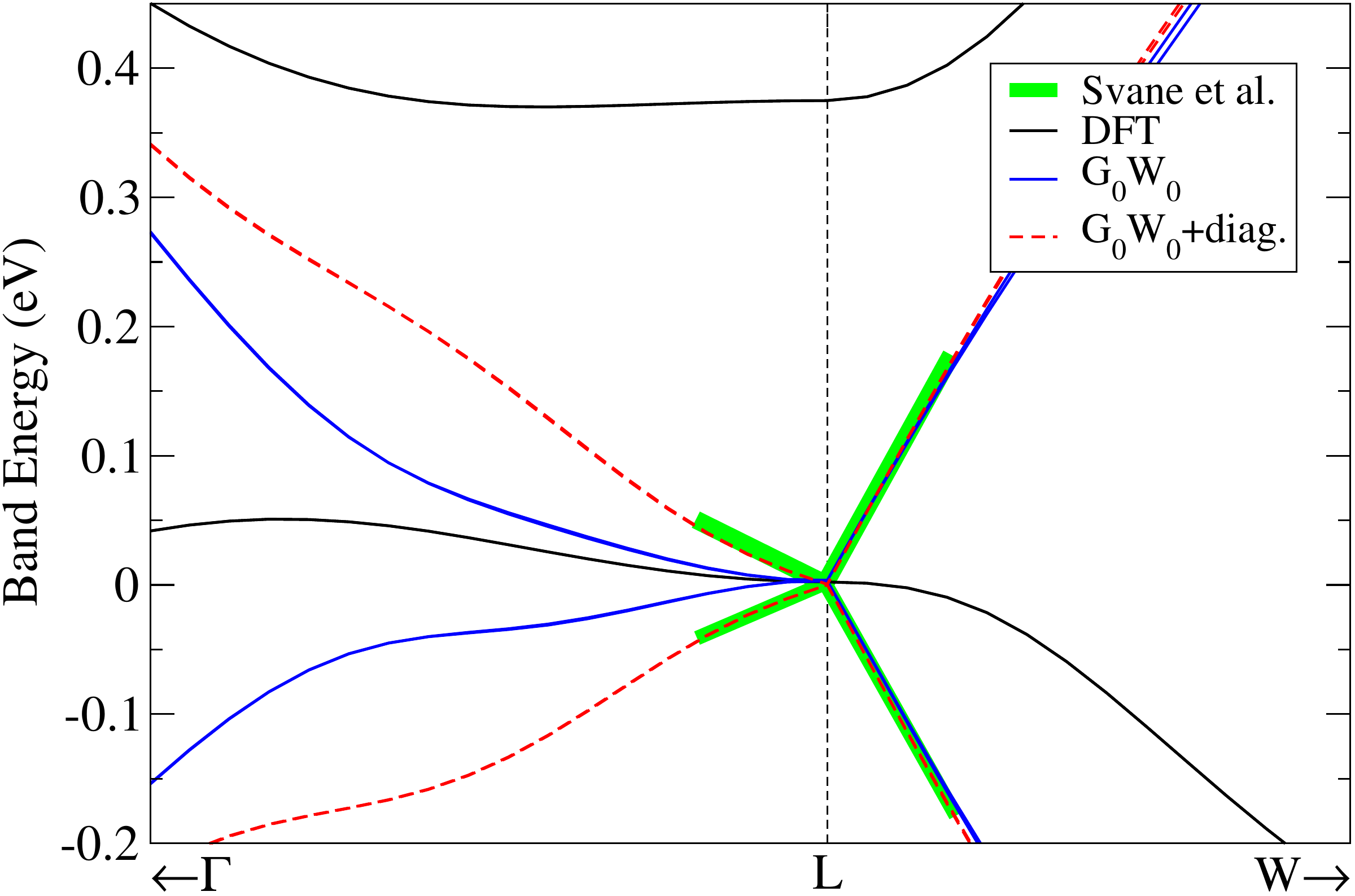}
        \caption{Band structure for compressed PbTe at the topological transition volume (9\% compression). Figure depicts the band structures obtained with DFT (black solid lines), diagonal $G_0W_0$ (blue solid) and $G_0W_0$ including off-diagonal corrections (red dashed). Results are compared with those obtained with quasiparticle self consistent $GW$ \cite{Svane2010a}.}
        \label{fig:dirac}
    \end{center}
\end{figure}

We show in Fig. \ref{fig:dirac} the band structure of PbTe at the volume of the transition from trivial to non-trivial topology (as obtained with $G_0W_0$). At the 9\% compressed volume, DFT predicts a topological material well past the transition point, with a relatively large inverted gap of around 0.4 eV. Instead, the $G_0W_0$ band structure shows a vanishing gap. Nevertheless, Fig. \ref{fig:dirac} shows that obtaining the linear dispersion characteristic of the topological transition requires the diagonalization of the $G_0W_0$ Hamiltonian. After taking into account the off-diagonal contribution we obtain a band structure in good agreement with the more sophisticated -- and computationally demanding -- quasiparticle self-consistent $GW$ \cite{Svane2010a}.

Finally, for GeTe the situation is similar to PbTe, with the exception that for this material both DFT and $G_0W_0$ give the same qualitative result, i.e. a band crossover at positive pressures.  Using the band structure as a function of volume calculated with $G_0W_0$ we obtain a critical pressure of 1.9 GPa. DFT overestimates the critical pressure, yielding a value of 3.0 GPa. 

\begin{table}[]
    \begin{tabularx}{\columnwidth}{X|XXXX}
    \hline \hline
            & $K$ (GPa) & $K'$ & $a_\mathrm{eq}$ (\AA) & $a_\mathrm{exp}$ (\AA) \\
    \hline
        PbTe & 43.6 & 4.6 & 6.43 & 6.46 ~\cite{MadelungHandbook}\\
        SnTe & 40.0 & 4.5 & 6.42 & 6.33 ~\cite{MadelungHandbook}\\
        GeTe & 43.5 & 4.2 & 6.07 & 6.01 ~\cite{Wiedemeier1977}\\
    \hline \hline
    \end{tabularx}
    \caption{Structural parameters obtained from fits to a Murnaghan equation of state, used to obtain the pressures in Table \ref{tab:transitionPressures} from the transition volumes. The listed parameters are the bulk modulus $K$, the linear coefficient of the expansion $K(P)$, $K'$, the equilibrium lattice parameter obtained from the fits and the experimental lattice parameter used to obtain the band structure presented in Sec. \ref{sec:electronic struct exp}. }
    \label{tab:Murnaghan}
\end{table}

\begin{table}[]
    \begin{tabularx}{\columnwidth}{X|XXX}
    \hline \hline 
            & \multicolumn{3}{c}{$\partial E_g/ \partial \ln{V}$ (eV)} \\
    \hline
            & PbTe & SnTe & GeTe \\
    \hline
        DFT        & -2.3       & -3.08 & 3.22 \\
        $G_0W_0$   &  2.8       & -3.51 & 3.92 \\
        Exp.       &  2.9 - 3.0 & --    & -- \\
    \hline \hline
    \end{tabularx}
    \caption{Deformation potentials in eV for GeTe, SnTe and PbTe, calcualted both with DFT and $G_0W_0$. Experimental data for PbTe from Ref. \onlinecite{Nimtz1983,Zasavitskii2004}. }
    \label{tab:defPot}
\end{table}

Data in Fig. \ref{fig:gapVsAlat} can also be used to obtain the deformation potential $\partial E_g/\partial\ln{ V}$, collected in Table \ref{tab:defPot}. 
For PbTe we can see that, in addition to correcting the spurious negative sign obtained within DFT, $G_0W_0$ accurately reproduces the experimental value. 
Similar results were obtained before for PbTe using hybrid DFT \cite{Murphy2018a}.
For either GeTe or SnTe we were not able to find previous reports of this coefficient, but since the ordering of the bands obtained within DFT is the correct one we observe that the correction from $G_0W_0$ is only quantitative. For all three materials we find that $G_0W_0$ tends to increase the absolute value of the deformation potential with respect to the DFT result.

\section{Discussion}
Since we are suggesting here the prospect of a topological transition with pressure of the electronic structure for both PbTe and GeTe, the possibility of structural phase transitions that may interfere with the band crossover deserves some discussion.
According to synchrotron x-ray diffraction experiments, PbTe presents a phase transition from the cubic rocksalt phase to orthorombic $Pnma$ at 6.7 GPa \cite{Li2013e,Rousse2005}. Therefore, according to our results the transition from a trivial to a non-trivial topology of the band structure would occur before the structural phase transition.

The case of GeTe is more complex, since at ambient conditions its structure presents an rombohedral distortion with respect to the high temperature cubic phase analyzed in this work. The rombohedral distortion consists in a relative displacement of the two sublattices along the pseudocubic [111] direction, accompanied with an elongation of the cell along the same direction. High pressure Raman scattering measurements show that the off-centering disappears at $\sim 3$ GPa, and the complete transition to cubic occurs for $P< 6$ GPa~\cite{Pawbake2019}. 
Since our constrained calculations for the cubic phase predict the electronic topological transition to occur at a lower pressure than the structural one, we expect GeTe to be a crystalline topological insulator for pressures higher than the structural transition pressure of $\sim 6$ GPa.

\section{Conclusions}
We have performed a comparative study of the $G_0W_0$ corrections, including off-diagonal contributions to the self energy, to the DFT band structure of PbTe, SnTe and GeTe, three narrow-gap semiconductors with a band structure sitting very close to the crossover between a trivial and a non-trivial topology. We have shown that, for PbTe, the conventional diagonal-only $G_0W_0$ produces artifacts near the band gap edges due to the inability of the method to 
disentangle the DFT band structure ---which presents a wrong ordering at L. These artifacts disappear when the $G_0W_0$ corrections are calculated for the subspace of the entangled DFT states, including off-diagonal contributions, and the quasiparticle energy is obtained by diagonalization of the resulting matrix. 

The evolution of the gap with the volume of the unit cell reveals that both PbTe and cubic GeTe may undergo a transition from a narrow gap semiconductor with a trivial topology of the band structure into a topological insulator. In the case of PbTe this transition occurs at around 4.8 GPa, below the pressure for the first structural phase transition into a rombohedral phase. GeTe crystallizes in the cubic phase for pressures above 6 GPa, higher than the critical pressure for the band crossover, and therefore should display topological insulator characteristics. 

\section{Acknowledgements}
This work was funded by the Science Foundation Ireland and the Department of Economy (Northern Ireland) under investigators Program, grant 15/IA/3160. Simulations were carried out on computational facilities from Queen's University Belfast. Additional computational support was provided by the UK Materials and Molecular Modelling Hub, which is partially funded by EPSRC (EP/P020194).

\appendix 
\section{Quasi-degenerate $G_0W_0$ approach}\label{sc:app}
The eigenvalues of Eq.~\eqref{eq:offdiag}
cannot be obtained straightforwardly since (i) the set of equations is non-linear; (ii) $H^{GW}_{nm}$ is non-Hermitian and, in general, does not have real eigenvalues.

As a remedy to the latter issue, we propose a $G_0W_0$-like approach which works when there are few non-negligible $\Delta \Sigma_{\bm{k}nm} = \mel{\phi^\mathrm{DFT}_{\bm{k}n}}{\Sigma(E^\mathrm{QP}_{\bm{k}n})  - V_{xc}}{\phi^\mathrm{DFT}_{\bm{k}m}}$ and the corresponding KS states are quasidegenerate. 

Let's call $\cal D$ the manifold of quasidegenerate KS states and define (in what follows, we drop the $\bm{k}$ index for simplicity):
\begin{equation}
    \hat \Delta = \sum_{i \in \cal D} | \phi_i \rangle (\bar \varepsilon - \varepsilon_i )\langle \phi_i |,
\end{equation}
as the scissor operator that makes all $i\in \cal D$ degenerate with an energy $\bar \varepsilon$.
We define a new independent-particle system with a Hamiltonian $$\hat h^s = \hat h + \hat \Delta,$$ where $\hat h$ is the KS Hamiltonian. The many-body perturbation to that system is $\hat \Sigma (\omega) - \hat v_{\rm xc} - \hat \Delta$.    

 Provided that $\bar \varepsilon - \varepsilon_i$ are small, perturbation theory should provide the same results (within the desired accuracy) starting either with $\hat h$ and $\hat h^s$. We choose a convenient definition of $\bar \varepsilon$ as the average $\left(\sum_{i\in \cal D} \varepsilon_i\right)/\dim\cal D$. 
 For $\hat h_s$, we can apply degenerate perturbation theory to $\cal D$, 
\begin{align}\label{eq:qdpt}
\left({\bf \Delta \Sigma}(\omega)-\delta\omega{\bf I} \right ) {\bf c} = 0,
\end{align}
where $\bf \delta\omega$ is $\omega -\bar\varepsilon$ and $\bf c$ is the vector of overlaps between the KS states in $\cal D$ and the corresponding QP states, $\langle \phi_n|\psi_l\rangle$. 
 Next, we linearize Eq.~\eqref{eq:qdpt} similarly to what is customary in the diagonal-only case \cite{Aryasetiawan1998} and we get
\begin{equation}\label{eq:leneq}
\left({\bf \Delta\bar \Sigma} -\delta\omega {\bf C} \right ) {\bf c} = 0.    
\end{equation}
This is a generalized eigenvalue problem in which $\bf \Delta \bar\Sigma$ is the matrix of $\Delta \Sigma_{mn}(\bar \varepsilon)$ and $ \bf $ the matrix of $C_{mn} = \delta_{mn} - \partial_\omega \Sigma_{mn}|_{\omega=\bar \varepsilon}$
(please note the relation between $C_{nn}$ and the renormalization factor, $Z_n$, usually introduced in the linearization of the diagonal-only case, $Z_n = C_{nn}^{-1}$ ~\cite{Aryasetiawan1998}) 
Both matrices are \emph{Hermitian} and Eq.~\eqref{eq:leneq} can be solved by Schur decomposition:
\begin{equation}
{\bf Q}({\bf \Delta\bar \Sigma} - \delta\omega {\bf C}){\bf Q} = ({\bf S} - \delta\omega {\bf T})    
\end{equation}
where ${\bf Q}$ is orthogonal and ${\bf S}$ and ${\bf T}$ are upper triangular matrices which have the same eigenvalues as $\bf \Delta\bar \Sigma$ and $\bf C$, respectively. From the properties of triangular matrices we can then find the quasiparticle corrections,
\begin{equation}
\delta\omega_l = T_{nn}^{-1} S_{nn},
\end{equation}
which are ensured to be real.

For the case of PbTe we simplified the above approach. 
First, 
we avoid the direct calculation of the self-energy matrix elements at $\bar \varepsilon$. For the off-diagonal elements we observe that for a pair of quasidegenerate states $i$ and $j$, 
\begin{equation}
\Sigma_{ij}\left(\frac{\varepsilon_i  +\varepsilon_j}{2}\right) \approx \frac{\Sigma_{ij}(\varepsilon_i)+\Sigma_{ij}(\varepsilon_j)}{2}.\label{eq:avsigma}
\end{equation}
For the diagonal elements, $\Sigma_{ii}\left(\bar \varepsilon \right) \approx \Sigma_{ii}\left(\varepsilon_i \right)$.
Second, 
we observed that the off-diagonal elements of $\bf C$ are negligible, i.e $C_{mn} \approx C_{mm}\delta_{mn}$ and therefore $T_{mm}^{-1}\approx C_{mm}^{-1}=Z_m$. Then, we first obtained the eigenvalues of $\bf \Delta \bar \Sigma$, i.e. $\lambda_m = \bar E_m -\bar\varepsilon$, and multiply them by $Z_m$.

At $L$, we have a manifold of 4 quasidegenerate KS states, 2 spin-up and 2 spin-down. Then, we have two $2\times 2$ eigenproblems for which we evaluate $\Sigma(\bar\varepsilon)$ as above. Finally, we also considered the original eigenvalue problem and verified that the quasiparticle corrections have a very small imaginary part while the real part is very close to the eigenvalues from the hermitized eigenproblem.



\begin{thebibliography}{51}%
\makeatletter
\providecommand \@ifxundefined [1]{%
 \@ifx{#1\undefined}
}%
\providecommand \@ifnum [1]{%
 \ifnum #1\expandafter \@firstoftwo
 \else \expandafter \@secondoftwo
 \fi
}%
\providecommand \@ifx [1]{%
 \ifx #1\expandafter \@firstoftwo
 \else \expandafter \@secondoftwo
 \fi
}%
\providecommand \natexlab [1]{#1}%
\providecommand \enquote  [1]{``#1''}%
\providecommand \bibnamefont  [1]{#1}%
\providecommand \bibfnamefont [1]{#1}%
\providecommand \citenamefont [1]{#1}%
\providecommand \href@noop [0]{\@secondoftwo}%
\providecommand \href [0]{\begingroup \@sanitize@url \@href}%
\providecommand \@href[1]{\@@startlink{#1}\@@href}%
\providecommand \@@href[1]{\endgroup#1\@@endlink}%
\providecommand \@sanitize@url [0]{\catcode `\\12\catcode `\$12\catcode
  `\&12\catcode `\#12\catcode `\^12\catcode `\_12\catcode `\%12\relax}%
\providecommand \@@startlink[1]{}%
\providecommand \@@endlink[0]{}%
\providecommand \url  [0]{\begingroup\@sanitize@url \@url }%
\providecommand \@url [1]{\endgroup\@href {#1}{\urlprefix }}%
\providecommand \urlprefix  [0]{URL }%
\providecommand \Eprint [0]{\href }%
\providecommand \doibase [0]{http://dx.doi.org/}%
\providecommand \selectlanguage [0]{\@gobble}%
\providecommand \bibinfo  [0]{\@secondoftwo}%
\providecommand \bibfield  [0]{\@secondoftwo}%
\providecommand \translation [1]{[#1]}%
\providecommand \BibitemOpen [0]{}%
\providecommand \bibitemStop [0]{}%
\providecommand \bibitemNoStop [0]{.\EOS\space}%
\providecommand \EOS [0]{\spacefactor3000\relax}%
\providecommand \BibitemShut  [1]{\csname bibitem#1\endcsname}%
\let\auto@bib@innerbib\@empty
\bibitem [{\citenamefont {Rosi}(1968)}]{Rosi1968}%
  \BibitemOpen
  \bibfield  {author} {\bibinfo {author} {\bibfnamefont {F.}~\bibnamefont
  {Rosi}},\ }\href {\doibase 10.1016/0038-1101(68)90104-4} {\bibfield
  {journal} {\bibinfo  {journal} {Solid. State. Electron.}\ }\textbf {\bibinfo
  {volume} {11}},\ \bibinfo {pages} {833} (\bibinfo {year} {1968})}\BibitemShut
  {NoStop}%
\bibitem [{\citenamefont {Wood}(1988)}]{Wood1988}%
  \BibitemOpen
  \bibfield  {author} {\bibinfo {author} {\bibfnamefont {C.}~\bibnamefont
  {Wood}},\ }\href {\doibase 10.1088/0034-4885/51/4/001} {\bibfield  {journal}
  {\bibinfo  {journal} {Rep. Prog. Phys.}\ }\textbf {\bibinfo {volume} {51}},\
  \bibinfo {pages} {459} (\bibinfo {year} {1988})}\BibitemShut {NoStop}%
\bibitem [{\citenamefont {Snyder}\ and\ \citenamefont
  {Toberer}(2008)}]{Snyder2008}%
  \BibitemOpen
  \bibfield  {author} {\bibinfo {author} {\bibfnamefont {G.~J.}\ \bibnamefont
  {Snyder}}\ and\ \bibinfo {author} {\bibfnamefont {E.~S.}\ \bibnamefont
  {Toberer}},\ }\href@noop {} {\bibfield  {journal} {\bibinfo  {journal} {Nat.
  Mater.}\ }\textbf {\bibinfo {volume} {7}},\ \bibinfo {pages} {105} (\bibinfo
  {year} {2008})}\BibitemShut {NoStop}%
\bibitem [{\citenamefont {Kolobov}\ \emph {et~al.}(2004)\citenamefont
  {Kolobov}, \citenamefont {Fons}, \citenamefont {Frenkel}, \citenamefont
  {Ankudinov}, \citenamefont {Tominaga},\ and\ \citenamefont
  {Uruga}}]{Kolobov2004}%
  \BibitemOpen
  \bibfield  {author} {\bibinfo {author} {\bibfnamefont {A.~V.}\ \bibnamefont
  {Kolobov}}, \bibinfo {author} {\bibfnamefont {P.}~\bibnamefont {Fons}},
  \bibinfo {author} {\bibfnamefont {A.~I.}\ \bibnamefont {Frenkel}}, \bibinfo
  {author} {\bibfnamefont {A.~L.}\ \bibnamefont {Ankudinov}}, \bibinfo {author}
  {\bibfnamefont {J.}~\bibnamefont {Tominaga}}, \ and\ \bibinfo {author}
  {\bibfnamefont {T.}~\bibnamefont {Uruga}},\ }\href {\doibase
  10.1038/nmat1215} {\bibfield  {journal} {\bibinfo  {journal} {Nat. Mater.}\
  }\textbf {\bibinfo {volume} {3}},\ \bibinfo {pages} {703} (\bibinfo {year}
  {2004})}\BibitemShut {NoStop}%
\bibitem [{\citenamefont {Schneider}\ \emph {et~al.}(2010)\citenamefont
  {Schneider}, \citenamefont {Rosenthal}, \citenamefont {Stiewe},\ and\
  \citenamefont {Oeckler}}]{Schneider2010}%
  \BibitemOpen
  \bibfield  {author} {\bibinfo {author} {\bibfnamefont {M.~N.}\ \bibnamefont
  {Schneider}}, \bibinfo {author} {\bibfnamefont {T.}~\bibnamefont
  {Rosenthal}}, \bibinfo {author} {\bibfnamefont {C.}~\bibnamefont {Stiewe}}, \
  and\ \bibinfo {author} {\bibfnamefont {O.}~\bibnamefont {Oeckler}},\ }\href
  {\doibase 10.1524/zkri.2010.1320} {\bibfield  {journal} {\bibinfo  {journal}
  {Zeitschrift fur Krist.}\ }\textbf {\bibinfo {volume} {225}},\ \bibinfo
  {pages} {463} (\bibinfo {year} {2010})}\BibitemShut {NoStop}%
\bibitem [{\citenamefont {Hsieh}\ \emph {et~al.}(2012)\citenamefont {Hsieh},
  \citenamefont {Lin}, \citenamefont {Liu}, \citenamefont {Duan}, \citenamefont
  {Bansil},\ and\ \citenamefont {Fu}}]{Hsieh2012}%
  \BibitemOpen
  \bibfield  {author} {\bibinfo {author} {\bibfnamefont {T.~H.}\ \bibnamefont
  {Hsieh}}, \bibinfo {author} {\bibfnamefont {H.}~\bibnamefont {Lin}}, \bibinfo
  {author} {\bibfnamefont {J.}~\bibnamefont {Liu}}, \bibinfo {author}
  {\bibfnamefont {W.}~\bibnamefont {Duan}}, \bibinfo {author} {\bibfnamefont
  {A.}~\bibnamefont {Bansil}}, \ and\ \bibinfo {author} {\bibfnamefont
  {L.}~\bibnamefont {Fu}},\ }\href {\doibase 10.1038/ncomms1969} {\bibfield
  {journal} {\bibinfo  {journal} {Nat. Comms.}\ }\textbf {\bibinfo {volume}
  {3}},\ \bibinfo {pages} {982} (\bibinfo {year} {2012})}\BibitemShut {NoStop}%
\bibitem [{\citenamefont {Tanaka}\ \emph {et~al.}(2012)\citenamefont {Tanaka},
  \citenamefont {Ren}, \citenamefont {Sato}, \citenamefont {Nakayama},
  \citenamefont {Souma}, \citenamefont {Takahashi}, \citenamefont {Segawa},\
  and\ \citenamefont {Ando}}]{Tanaka2012}%
  \BibitemOpen
  \bibfield  {author} {\bibinfo {author} {\bibfnamefont {Y.}~\bibnamefont
  {Tanaka}}, \bibinfo {author} {\bibfnamefont {Z.}~\bibnamefont {Ren}},
  \bibinfo {author} {\bibfnamefont {T.}~\bibnamefont {Sato}}, \bibinfo {author}
  {\bibfnamefont {K.}~\bibnamefont {Nakayama}}, \bibinfo {author}
  {\bibfnamefont {S.}~\bibnamefont {Souma}}, \bibinfo {author} {\bibfnamefont
  {T.}~\bibnamefont {Takahashi}}, \bibinfo {author} {\bibfnamefont
  {K.}~\bibnamefont {Segawa}}, \ and\ \bibinfo {author} {\bibfnamefont
  {Y.}~\bibnamefont {Ando}},\ }\href {\doibase 10.1038/nphys2442} {\bibfield
  {journal} {\bibinfo  {journal} {Nat. Phys.}\ }\textbf {\bibinfo {volume}
  {8}},\ \bibinfo {pages} {800} (\bibinfo {year} {2012})}\BibitemShut {NoStop}%
\bibitem [{\citenamefont {Fu}(2011)}]{Fu2011}%
  \BibitemOpen
  \bibfield  {author} {\bibinfo {author} {\bibfnamefont {L.}~\bibnamefont
  {Fu}},\ }\href {\doibase 10.1103/PhysRevLett.106.106802} {\bibfield
  {journal} {\bibinfo  {journal} {Phys. Rev. Lett.}\ }\textbf {\bibinfo
  {volume} {106}},\ \bibinfo {pages} {1} (\bibinfo {year} {2011})},\ \Eprint
  {http://arxiv.org/abs/1010.1802} {arXiv:1010.1802} \BibitemShut {NoStop}%
\bibitem [{\citenamefont {Ye}\ \emph {et~al.}(2015)\citenamefont {Ye},
  \citenamefont {Deng}, \citenamefont {Wu}, \citenamefont {Li}, \citenamefont
  {Wei},\ and\ \citenamefont {Luo}}]{Ye2015}%
  \BibitemOpen
  \bibfield  {author} {\bibinfo {author} {\bibfnamefont {Z.-Y.}\ \bibnamefont
  {Ye}}, \bibinfo {author} {\bibfnamefont {H.-X.}\ \bibnamefont {Deng}},
  \bibinfo {author} {\bibfnamefont {H.-Z.}\ \bibnamefont {Wu}}, \bibinfo
  {author} {\bibfnamefont {S.-S.}\ \bibnamefont {Li}}, \bibinfo {author}
  {\bibfnamefont {S.-H.}\ \bibnamefont {Wei}}, \ and\ \bibinfo {author}
  {\bibfnamefont {J.-W.}\ \bibnamefont {Luo}},\ }\href {\doibase
  10.1038/npjcompumats.2015.1} {\bibfield  {journal} {\bibinfo  {journal} {npj
  Comput. Mater.}\ }\textbf {\bibinfo {volume} {1}},\ \bibinfo {pages} {15001}
  (\bibinfo {year} {2015})}\BibitemShut {NoStop}%
\bibitem [{\citenamefont {Hummer}\ \emph {et~al.}(2007)\citenamefont {Hummer},
  \citenamefont {Gr{\"{u}}neis},\ and\ \citenamefont {Kresse}}]{Hummer2007}%
  \BibitemOpen
  \bibfield  {author} {\bibinfo {author} {\bibfnamefont {K.}~\bibnamefont
  {Hummer}}, \bibinfo {author} {\bibfnamefont {A.}~\bibnamefont
  {Gr{\"{u}}neis}}, \ and\ \bibinfo {author} {\bibfnamefont {G.}~\bibnamefont
  {Kresse}},\ }\href {\doibase 10.1103/PhysRevB.75.195211} {\bibfield
  {journal} {\bibinfo  {journal} {Phys. Rev. B}\ }\textbf {\bibinfo {volume}
  {75}},\ \bibinfo {pages} {195211} (\bibinfo {year} {2007})}\BibitemShut
  {NoStop}%
\bibitem [{\citenamefont {Svane}\ \emph {et~al.}(2010)\citenamefont {Svane},
  \citenamefont {Christensen}, \citenamefont {Cardona}, \citenamefont
  {Chantis}, \citenamefont {{Van Schilfgaarde}},\ and\ \citenamefont
  {Kotani}}]{Svane2010a}%
  \BibitemOpen
  \bibfield  {author} {\bibinfo {author} {\bibfnamefont {A.}~\bibnamefont
  {Svane}}, \bibinfo {author} {\bibfnamefont {N.~E.}\ \bibnamefont
  {Christensen}}, \bibinfo {author} {\bibfnamefont {M.}~\bibnamefont
  {Cardona}}, \bibinfo {author} {\bibfnamefont {A.~N.}\ \bibnamefont
  {Chantis}}, \bibinfo {author} {\bibfnamefont {M.}~\bibnamefont {{Van
  Schilfgaarde}}}, \ and\ \bibinfo {author} {\bibfnamefont {T.}~\bibnamefont
  {Kotani}},\ }\href {\doibase 10.1103/PhysRevB.81.245120} {\bibfield
  {journal} {\bibinfo  {journal} {Phys. Rev. B}\ }\textbf {\bibinfo {volume}
  {81}},\ \bibinfo {pages} {245120} (\bibinfo {year} {2010})}\BibitemShut
  {NoStop}%
\bibitem [{\citenamefont {Robredo}\ \emph {et~al.}(2019)\citenamefont
  {Robredo}, \citenamefont {Vergniory},\ and\ \citenamefont
  {Bradlyn}}]{Robredo2019}%
  \BibitemOpen
  \bibfield  {author} {\bibinfo {author} {\bibfnamefont {I.}~\bibnamefont
  {Robredo}}, \bibinfo {author} {\bibfnamefont {M.~G.}\ \bibnamefont
  {Vergniory}}, \ and\ \bibinfo {author} {\bibfnamefont {B.}~\bibnamefont
  {Bradlyn}},\ }\href {\doibase 10.1103/PhysRevMaterials.3.041202} {\bibfield
  {journal} {\bibinfo  {journal} {Phys. Rev. Mater.}\ }\textbf {\bibinfo
  {volume} {3}},\ \bibinfo {pages} {041202(R)} (\bibinfo {year}
  {2019})}\BibitemShut {NoStop}%
\bibitem [{\citenamefont {Xu}\ \emph {et~al.}(2012)\citenamefont {Xu},
  \citenamefont {Liu}, \citenamefont {Alidoust}, \citenamefont {Neupane},
  \citenamefont {Qian}, \citenamefont {Belopolski}, \citenamefont {Denlinger},
  \citenamefont {Wang}, \citenamefont {Lin}, \citenamefont {Wray},
  \citenamefont {Landolt}, \citenamefont {Slomski}, \citenamefont {Dil},
  \citenamefont {Marcinkova}, \citenamefont {Morosan}, \citenamefont {Gibson},
  \citenamefont {Sankar}, \citenamefont {Chou}, \citenamefont {Cava},
  \citenamefont {Bansil},\ and\ \citenamefont {Hasan}}]{Xu2012}%
  \BibitemOpen
  \bibfield  {author} {\bibinfo {author} {\bibfnamefont {S.~Y.}\ \bibnamefont
  {Xu}}, \bibinfo {author} {\bibfnamefont {C.}~\bibnamefont {Liu}}, \bibinfo
  {author} {\bibfnamefont {N.}~\bibnamefont {Alidoust}}, \bibinfo {author}
  {\bibfnamefont {M.}~\bibnamefont {Neupane}}, \bibinfo {author} {\bibfnamefont
  {D.}~\bibnamefont {Qian}}, \bibinfo {author} {\bibfnamefont {I.}~\bibnamefont
  {Belopolski}}, \bibinfo {author} {\bibfnamefont {J.~D.}\ \bibnamefont
  {Denlinger}}, \bibinfo {author} {\bibfnamefont {Y.~J.}\ \bibnamefont {Wang}},
  \bibinfo {author} {\bibfnamefont {H.}~\bibnamefont {Lin}}, \bibinfo {author}
  {\bibfnamefont {L.~A.}\ \bibnamefont {Wray}}, \bibinfo {author}
  {\bibfnamefont {G.}~\bibnamefont {Landolt}}, \bibinfo {author} {\bibfnamefont
  {B.}~\bibnamefont {Slomski}}, \bibinfo {author} {\bibfnamefont {J.~H.}\
  \bibnamefont {Dil}}, \bibinfo {author} {\bibfnamefont {A.}~\bibnamefont
  {Marcinkova}}, \bibinfo {author} {\bibfnamefont {E.}~\bibnamefont {Morosan}},
  \bibinfo {author} {\bibfnamefont {Q.}~\bibnamefont {Gibson}}, \bibinfo
  {author} {\bibfnamefont {R.}~\bibnamefont {Sankar}}, \bibinfo {author}
  {\bibfnamefont {F.~C.}\ \bibnamefont {Chou}}, \bibinfo {author}
  {\bibfnamefont {R.~J.}\ \bibnamefont {Cava}}, \bibinfo {author}
  {\bibfnamefont {A.}~\bibnamefont {Bansil}}, \ and\ \bibinfo {author}
  {\bibfnamefont {M.~Z.}\ \bibnamefont {Hasan}},\ }\href {\doibase
  10.1038/ncomms2191} {\bibfield  {journal} {\bibinfo  {journal} {Nat.
  Commun.}\ }\textbf {\bibinfo {volume} {3}},\ \bibinfo {pages} {1192}
  (\bibinfo {year} {2012})}\BibitemShut {NoStop}%
\bibitem [{\citenamefont {Yan}\ \emph {et~al.}(2014)\citenamefont {Yan},
  \citenamefont {Liu}, \citenamefont {Zang}, \citenamefont {Wang},
  \citenamefont {Wang}, \citenamefont {Wang}, \citenamefont {Zhang},
  \citenamefont {Wang}, \citenamefont {Ma}, \citenamefont {Ji}, \citenamefont
  {He}, \citenamefont {Fu}, \citenamefont {Duan}, \citenamefont {Xue},\ and\
  \citenamefont {Chen}}]{Yan2014}%
  \BibitemOpen
  \bibfield  {author} {\bibinfo {author} {\bibfnamefont {C.}~\bibnamefont
  {Yan}}, \bibinfo {author} {\bibfnamefont {J.}~\bibnamefont {Liu}}, \bibinfo
  {author} {\bibfnamefont {Y.}~\bibnamefont {Zang}}, \bibinfo {author}
  {\bibfnamefont {J.}~\bibnamefont {Wang}}, \bibinfo {author} {\bibfnamefont
  {Z.}~\bibnamefont {Wang}}, \bibinfo {author} {\bibfnamefont {P.}~\bibnamefont
  {Wang}}, \bibinfo {author} {\bibfnamefont {Z.~D.}\ \bibnamefont {Zhang}},
  \bibinfo {author} {\bibfnamefont {L.}~\bibnamefont {Wang}}, \bibinfo {author}
  {\bibfnamefont {X.}~\bibnamefont {Ma}}, \bibinfo {author} {\bibfnamefont
  {S.}~\bibnamefont {Ji}}, \bibinfo {author} {\bibfnamefont {K.}~\bibnamefont
  {He}}, \bibinfo {author} {\bibfnamefont {L.}~\bibnamefont {Fu}}, \bibinfo
  {author} {\bibfnamefont {W.}~\bibnamefont {Duan}}, \bibinfo {author}
  {\bibfnamefont {Q.~K.}\ \bibnamefont {Xue}}, \ and\ \bibinfo {author}
  {\bibfnamefont {X.}~\bibnamefont {Chen}},\ }\href {\doibase
  10.1103/PhysRevLett.112.186801} {\bibfield  {journal} {\bibinfo  {journal}
  {Phys. Rev. Lett.}\ }\textbf {\bibinfo {volume} {112}},\ \bibinfo {pages} {1}
  (\bibinfo {year} {2014})}\BibitemShut {NoStop}%
\bibitem [{\citenamefont {Dimmock}\ \emph {et~al.}(1966)\citenamefont
  {Dimmock}, \citenamefont {Melngailis},\ and\ \citenamefont
  {Strauss}}]{Dimmock1966}%
  \BibitemOpen
  \bibfield  {author} {\bibinfo {author} {\bibfnamefont {J.~O.}\ \bibnamefont
  {Dimmock}}, \bibinfo {author} {\bibfnamefont {I.}~\bibnamefont {Melngailis}},
  \ and\ \bibinfo {author} {\bibfnamefont {A.~J.}\ \bibnamefont {Strauss}},\
  }\href {http://link.aps.org/doi/10.1103/PhysRevLett.16.1193} {\bibfield
  {journal} {\bibinfo  {journal} {Phys. Rev. Lett.}\ }\textbf {\bibinfo
  {volume} {16}},\ \bibinfo {pages} {1193} (\bibinfo {year}
  {1966})}\BibitemShut {NoStop}%
\bibitem [{\citenamefont {Valeiko}\ \emph {et~al.}(1991)\citenamefont
  {Valeiko}, \citenamefont {Zasavitskii}, \citenamefont {Matveenko},\ and\
  \citenamefont {Matsonashvili}}]{Valeiko1991}%
  \BibitemOpen
  \bibfield  {author} {\bibinfo {author} {\bibfnamefont {M.~V.}\ \bibnamefont
  {Valeiko}}, \bibinfo {author} {\bibfnamefont {I.~I.}\ \bibnamefont
  {Zasavitskii}}, \bibinfo {author} {\bibfnamefont {A.~V.}\ \bibnamefont
  {Matveenko}}, \ and\ \bibinfo {author} {\bibfnamefont {B.~N.}\ \bibnamefont
  {Matsonashvili}},\ }\href@noop {} {\bibfield  {journal} {\bibinfo  {journal}
  {Superlattices Microstruct.}\ }\textbf {\bibinfo {volume} {9}},\ \bibinfo
  {pages} {195} (\bibinfo {year} {1991})}\BibitemShut {NoStop}%
\bibitem [{\citenamefont {Singh}(2013)}]{Singh2013a}%
  \BibitemOpen
  \bibfield  {author} {\bibinfo {author} {\bibfnamefont {D.~J.}\ \bibnamefont
  {Singh}},\ }\href {\doibase 10.1063/1.4807638} {\bibfield  {journal}
  {\bibinfo  {journal} {J. App. Phys.}\ }\textbf {\bibinfo {volume} {113}},\
  \bibinfo {pages} {203101} (\bibinfo {year} {2013})}\BibitemShut {NoStop}%
\bibitem [{\citenamefont {Murphy}\ \emph {et~al.}(2018)\citenamefont {Murphy},
  \citenamefont {Murphy-Armando}, \citenamefont {Fahy},\ and\ \citenamefont
  {Savi{\'{c}}}}]{Murphy2018a}%
  \BibitemOpen
  \bibfield  {author} {\bibinfo {author} {\bibfnamefont {A.~R.}\ \bibnamefont
  {Murphy}}, \bibinfo {author} {\bibfnamefont {F.}~\bibnamefont
  {Murphy-Armando}}, \bibinfo {author} {\bibfnamefont {S.}~\bibnamefont
  {Fahy}}, \ and\ \bibinfo {author} {\bibfnamefont {I.}~\bibnamefont
  {Savi{\'{c}}}},\ }\href {\doibase 10.1103/PhysRevB.98.085201} {\bibfield
  {journal} {\bibinfo  {journal} {Phys. Rev. B}\ }\textbf {\bibinfo {volume}
  {98}},\ \bibinfo {pages} {085201} (\bibinfo {year} {2018})}\BibitemShut
  {NoStop}%
\bibitem [{\citenamefont {Vidal}\ \emph {et~al.}(2011)\citenamefont {Vidal},
  \citenamefont {Zhang}, \citenamefont {Yu}, \citenamefont {Luo},\ and\
  \citenamefont {Zunger}}]{Vidal2011}%
  \BibitemOpen
  \bibfield  {author} {\bibinfo {author} {\bibfnamefont {J.}~\bibnamefont
  {Vidal}}, \bibinfo {author} {\bibfnamefont {X.}~\bibnamefont {Zhang}},
  \bibinfo {author} {\bibfnamefont {L.}~\bibnamefont {Yu}}, \bibinfo {author}
  {\bibfnamefont {J.~W.}\ \bibnamefont {Luo}}, \ and\ \bibinfo {author}
  {\bibfnamefont {A.}~\bibnamefont {Zunger}},\ }\href {\doibase
  10.1103/PhysRevB.84.041109} {\bibfield  {journal} {\bibinfo  {journal} {Phys.
  Rev. B}\ }\textbf {\bibinfo {volume} {84}},\ \bibinfo {pages} {1} (\bibinfo
  {year} {2011})}\BibitemShut {NoStop}%
\bibitem [{\citenamefont {{Van Schilfgaarde}}\ \emph
  {et~al.}(2006)\citenamefont {{Van Schilfgaarde}}, \citenamefont {Kotani},\
  and\ \citenamefont {Faleev}}]{VanSchilfgaarde2006}%
  \BibitemOpen
  \bibfield  {author} {\bibinfo {author} {\bibfnamefont {M.}~\bibnamefont {{Van
  Schilfgaarde}}}, \bibinfo {author} {\bibfnamefont {T.}~\bibnamefont
  {Kotani}}, \ and\ \bibinfo {author} {\bibfnamefont {S.~V.}\ \bibnamefont
  {Faleev}},\ }\href {\doibase 10.1103/PhysRevB.74.245125} {\bibfield
  {journal} {\bibinfo  {journal} {Phys. Rev. B}\ }\textbf {\bibinfo {volume}
  {74}},\ \bibinfo {pages} {245125} (\bibinfo {year} {2006})}\BibitemShut
  {NoStop}%
\bibitem [{\citenamefont {Aryasetiawan}\ and\ \citenamefont
  {Gunnarsson}(1998)}]{Aryasetiawan1998}%
  \BibitemOpen
  \bibfield  {author} {\bibinfo {author} {\bibfnamefont {F.}~\bibnamefont
  {Aryasetiawan}}\ and\ \bibinfo {author} {\bibfnamefont {O.}~\bibnamefont
  {Gunnarsson}},\ }\href {\doibase 10.1088/0034-4885/61/3/002} {\bibfield
  {journal} {\bibinfo  {journal} {Reports Prog. Phys.}\ }\textbf {\bibinfo
  {volume} {61}},\ \bibinfo {pages} {237} (\bibinfo {year} {1998})}\BibitemShut
  {NoStop}%
\bibitem [{\citenamefont {Aguilera}\ \emph {et~al.}(2013)\citenamefont
  {Aguilera}, \citenamefont {Friedrich}, \citenamefont {Bihlmayer},\ and\
  \citenamefont {Bl{\"{u}}gel}}]{Aguilera2013}%
  \BibitemOpen
  \bibfield  {author} {\bibinfo {author} {\bibfnamefont {I.}~\bibnamefont
  {Aguilera}}, \bibinfo {author} {\bibfnamefont {C.}~\bibnamefont {Friedrich}},
  \bibinfo {author} {\bibfnamefont {G.}~\bibnamefont {Bihlmayer}}, \ and\
  \bibinfo {author} {\bibfnamefont {S.}~\bibnamefont {Bl{\"{u}}gel}},\ }\href
  {\doibase 10.1103/PhysRevB.88.045206} {\bibfield  {journal} {\bibinfo
  {journal} {Phys. Rev. B}\ }\textbf {\bibinfo {volume} {88}},\ \bibinfo
  {pages} {045206} (\bibinfo {year} {2013})}\BibitemShut {NoStop}%
\bibitem [{\citenamefont {F{\"{o}}rster}\ \emph {et~al.}(2016)\citenamefont
  {F{\"{o}}rster}, \citenamefont {Kr{\"{u}}ger},\ and\ \citenamefont
  {Rohlfing}}]{Forster2016}%
  \BibitemOpen
  \bibfield  {author} {\bibinfo {author} {\bibfnamefont {T.}~\bibnamefont
  {F{\"{o}}rster}}, \bibinfo {author} {\bibfnamefont {P.}~\bibnamefont
  {Kr{\"{u}}ger}}, \ and\ \bibinfo {author} {\bibfnamefont {M.}~\bibnamefont
  {Rohlfing}},\ }\href {\doibase 10.1103/PhysRevB.93.205442} {\bibfield
  {journal} {\bibinfo  {journal} {Phys. Rev. B}\ }\textbf {\bibinfo {volume}
  {93}},\ \bibinfo {pages} {205442} (\bibinfo {year} {2016})}\BibitemShut
  {NoStop}%
\bibitem [{\citenamefont {Giannozzi}\ \emph {et~al.}(2009)\citenamefont
  {Giannozzi}, \citenamefont {Baroni}, \citenamefont {Bonini}, \citenamefont
  {Calandra}, \citenamefont {Car}, \citenamefont {Cavazzoni}, \citenamefont
  {Ceresoli}, \citenamefont {Chiarotti}, \citenamefont {Cococcioni},
  \citenamefont {Dabo}, \citenamefont {{Dal Corso}}, \citenamefont
  {de~Gironcoli}, \citenamefont {Fabris}, \citenamefont {Fratesi},
  \citenamefont {Gebauer}, \citenamefont {Gerstmann}, \citenamefont
  {Gougoussis}, \citenamefont {Kokalj}, \citenamefont {Lazzeri}, \citenamefont
  {Martin-Samos}, \citenamefont {Marzari}, \citenamefont {Mauri}, \citenamefont
  {Mazzarello}, \citenamefont {Paolini}, \citenamefont {Pasquarello},
  \citenamefont {Paulatto}, \citenamefont {Sbraccia}, \citenamefont {Scandolo},
  \citenamefont {Sclauzero}, \citenamefont {Seitsonen}, \citenamefont
  {Smogunov}, \citenamefont {Umari},\ and\ \citenamefont
  {Wentzcovitch}}]{Giannozzi2009}%
  \BibitemOpen
  \bibfield  {author} {\bibinfo {author} {\bibfnamefont {P.}~\bibnamefont
  {Giannozzi}}, \bibinfo {author} {\bibfnamefont {S.}~\bibnamefont {Baroni}},
  \bibinfo {author} {\bibfnamefont {N.}~\bibnamefont {Bonini}}, \bibinfo
  {author} {\bibfnamefont {M.}~\bibnamefont {Calandra}}, \bibinfo {author}
  {\bibfnamefont {R.}~\bibnamefont {Car}}, \bibinfo {author} {\bibfnamefont
  {C.}~\bibnamefont {Cavazzoni}}, \bibinfo {author} {\bibfnamefont
  {D.}~\bibnamefont {Ceresoli}}, \bibinfo {author} {\bibfnamefont {G.~L.}\
  \bibnamefont {Chiarotti}}, \bibinfo {author} {\bibfnamefont {M.}~\bibnamefont
  {Cococcioni}}, \bibinfo {author} {\bibfnamefont {I.}~\bibnamefont {Dabo}},
  \bibinfo {author} {\bibfnamefont {A.}~\bibnamefont {{Dal Corso}}}, \bibinfo
  {author} {\bibfnamefont {S.}~\bibnamefont {de~Gironcoli}}, \bibinfo {author}
  {\bibfnamefont {S.}~\bibnamefont {Fabris}}, \bibinfo {author} {\bibfnamefont
  {G.}~\bibnamefont {Fratesi}}, \bibinfo {author} {\bibfnamefont
  {R.}~\bibnamefont {Gebauer}}, \bibinfo {author} {\bibfnamefont
  {U.}~\bibnamefont {Gerstmann}}, \bibinfo {author} {\bibfnamefont
  {C.}~\bibnamefont {Gougoussis}}, \bibinfo {author} {\bibfnamefont
  {A.}~\bibnamefont {Kokalj}}, \bibinfo {author} {\bibfnamefont
  {M.}~\bibnamefont {Lazzeri}}, \bibinfo {author} {\bibfnamefont
  {L.}~\bibnamefont {Martin-Samos}}, \bibinfo {author} {\bibfnamefont
  {N.}~\bibnamefont {Marzari}}, \bibinfo {author} {\bibfnamefont
  {F.}~\bibnamefont {Mauri}}, \bibinfo {author} {\bibfnamefont
  {R.}~\bibnamefont {Mazzarello}}, \bibinfo {author} {\bibfnamefont
  {S.}~\bibnamefont {Paolini}}, \bibinfo {author} {\bibfnamefont
  {A.}~\bibnamefont {Pasquarello}}, \bibinfo {author} {\bibfnamefont
  {L.}~\bibnamefont {Paulatto}}, \bibinfo {author} {\bibfnamefont
  {C.}~\bibnamefont {Sbraccia}}, \bibinfo {author} {\bibfnamefont
  {S.}~\bibnamefont {Scandolo}}, \bibinfo {author} {\bibfnamefont
  {G.}~\bibnamefont {Sclauzero}}, \bibinfo {author} {\bibfnamefont {A.~P.}\
  \bibnamefont {Seitsonen}}, \bibinfo {author} {\bibfnamefont {A.}~\bibnamefont
  {Smogunov}}, \bibinfo {author} {\bibfnamefont {P.}~\bibnamefont {Umari}}, \
  and\ \bibinfo {author} {\bibfnamefont {R.~M.}\ \bibnamefont {Wentzcovitch}},\
  }\href {http://www.ncbi.nlm.nih.gov/pubmed/21832390} {\bibfield  {journal}
  {\bibinfo  {journal} {J. Phys. Condens. Matter}\ }\textbf {\bibinfo {volume}
  {21}},\ \bibinfo {pages} {395502} (\bibinfo {year} {2009})}\BibitemShut
  {NoStop}%
\bibitem [{\citenamefont {Giannozzi}\ \emph {et~al.}(2017)\citenamefont
  {Giannozzi}, \citenamefont {Andreussi}, \citenamefont {Brumme}, \citenamefont
  {Bunau}, \citenamefont {Nardelli}, \citenamefont {Calandra}, \citenamefont
  {Car}, \citenamefont {Cavazzoni}, \citenamefont {Ceresoli}, \citenamefont
  {Cococcioni},\ and\ \citenamefont {Others}}]{Giannozzi2017}%
  \BibitemOpen
  \bibfield  {author} {\bibinfo {author} {\bibfnamefont {P.}~\bibnamefont
  {Giannozzi}}, \bibinfo {author} {\bibfnamefont {O.}~\bibnamefont
  {Andreussi}}, \bibinfo {author} {\bibfnamefont {T.}~\bibnamefont {Brumme}},
  \bibinfo {author} {\bibfnamefont {O.}~\bibnamefont {Bunau}}, \bibinfo
  {author} {\bibfnamefont {M.~B.}\ \bibnamefont {Nardelli}}, \bibinfo {author}
  {\bibfnamefont {M.}~\bibnamefont {Calandra}}, \bibinfo {author}
  {\bibfnamefont {R.}~\bibnamefont {Car}}, \bibinfo {author} {\bibfnamefont
  {C.}~\bibnamefont {Cavazzoni}}, \bibinfo {author} {\bibfnamefont
  {D.}~\bibnamefont {Ceresoli}}, \bibinfo {author} {\bibfnamefont
  {M.}~\bibnamefont {Cococcioni}}, \ and\ \bibinfo {author} {\bibnamefont
  {Others}},\ }\href {http://arxiv.org/abs/1709.10010} {\bibfield  {journal}
  {\bibinfo  {journal} {J. Phys. Condens. Matterondensed Matter}\ }\textbf
  {\bibinfo {volume} {29}},\ \bibinfo {pages} {465901} (\bibinfo {year}
  {2017})}\BibitemShut {NoStop}%
\bibitem [{\citenamefont {Perdew}\ \emph {et~al.}(1996)\citenamefont {Perdew},
  \citenamefont {Burke},\ and\ \citenamefont {Ernzerhof}}]{Perdew1996}%
  \BibitemOpen
  \bibfield  {author} {\bibinfo {author} {\bibfnamefont {J.~P.}\ \bibnamefont
  {Perdew}}, \bibinfo {author} {\bibfnamefont {K.}~\bibnamefont {Burke}}, \
  and\ \bibinfo {author} {\bibfnamefont {M.}~\bibnamefont {Ernzerhof}},\ }\href
  {\doibase 10.1103/PhysRevLett.77.3865} {\bibfield  {journal} {\bibinfo
  {journal} {Phys. Rev. Lett.}\ }\textbf {\bibinfo {volume} {77}},\ \bibinfo
  {pages} {3865} (\bibinfo {year} {1996})}\BibitemShut {NoStop}%
\bibitem [{\citenamefont {van Setten}\ \emph {et~al.}(2018)\citenamefont {van
  Setten}, \citenamefont {Giantomassi}, \citenamefont {Bousquet}, \citenamefont
  {Verstraete}, \citenamefont {Hamann}, \citenamefont {Gonze},\ and\
  \citenamefont {Rignanese}}]{VanSetten2018}%
  \BibitemOpen
  \bibfield  {author} {\bibinfo {author} {\bibfnamefont {M.~J.}\ \bibnamefont
  {van Setten}}, \bibinfo {author} {\bibfnamefont {M.}~\bibnamefont
  {Giantomassi}}, \bibinfo {author} {\bibfnamefont {E.}~\bibnamefont
  {Bousquet}}, \bibinfo {author} {\bibfnamefont {M.~J.}\ \bibnamefont
  {Verstraete}}, \bibinfo {author} {\bibfnamefont {D.~R.}\ \bibnamefont
  {Hamann}}, \bibinfo {author} {\bibfnamefont {X.}~\bibnamefont {Gonze}}, \
  and\ \bibinfo {author} {\bibfnamefont {G.~M.}\ \bibnamefont {Rignanese}},\
  }\href {\doibase 10.1016/j.cpc.2018.01.012} {\bibfield  {journal} {\bibinfo
  {journal} {Comput. Phys. Commun.}\ }\textbf {\bibinfo {volume} {226}},\
  \bibinfo {pages} {39} (\bibinfo {year} {2018})}\BibitemShut {NoStop}%
\bibitem [{\citenamefont {Hamann}(2013)}]{Hamann2013}%
  \BibitemOpen
  \bibfield  {author} {\bibinfo {author} {\bibfnamefont {D.~R.}\ \bibnamefont
  {Hamann}},\ }\href {\doibase 10.1103/PhysRevB.88.085117} {\bibfield
  {journal} {\bibinfo  {journal} {Phys. Rev. B}\ }\textbf {\bibinfo {volume}
  {88}},\ \bibinfo {pages} {085117} (\bibinfo {year} {2013})}\BibitemShut
  {NoStop}%
\bibitem [{\citenamefont {Monkhorst}\ and\ \citenamefont
  {Pack}(1976)}]{Monkhorst1976}%
  \BibitemOpen
  \bibfield  {author} {\bibinfo {author} {\bibfnamefont {H.}~\bibnamefont
  {Monkhorst}}\ and\ \bibinfo {author} {\bibfnamefont {J.}~\bibnamefont
  {Pack}},\ }\href
  {http://www.if.pwr.wroc.pl/{~}scharoch/Abinitio/MonkhorstPack.pdf} {\bibfield
   {journal} {\bibinfo  {journal} {Phys. Rev. B}\ }\textbf {\bibinfo {volume}
  {13}},\ \bibinfo {pages} {5188} (\bibinfo {year} {1976})}\BibitemShut
  {NoStop}%
\bibitem [{\citenamefont {Marini}\ \emph {et~al.}(2009)\citenamefont {Marini},
  \citenamefont {Hogan}, \citenamefont {Gr{\"{u}}ning},\ and\ \citenamefont
  {Varsano}}]{Marini2009}%
  \BibitemOpen
  \bibfield  {author} {\bibinfo {author} {\bibfnamefont {A.}~\bibnamefont
  {Marini}}, \bibinfo {author} {\bibfnamefont {C.}~\bibnamefont {Hogan}},
  \bibinfo {author} {\bibfnamefont {M.}~\bibnamefont {Gr{\"{u}}ning}}, \ and\
  \bibinfo {author} {\bibfnamefont {D.}~\bibnamefont {Varsano}},\ }\href
  {\doibase 10.1016/j.cpc.2009.02.003} {\bibfield  {journal} {\bibinfo
  {journal} {Comput. Phys. Commun.}\ }\textbf {\bibinfo {volume} {180}},\
  \bibinfo {pages} {1392} (\bibinfo {year} {2009})}\BibitemShut {NoStop}%
\bibitem [{\citenamefont {Sangalli}\ \emph {et~al.}(2019)\citenamefont
  {Sangalli}, \citenamefont {Ferretti}, \citenamefont {Miranda}, \citenamefont
  {Attaccalite}, \citenamefont {Marri}, \citenamefont {Cannuccia},
  \citenamefont {Melo}, \citenamefont {Marsili}, \citenamefont {Paleari},
  \citenamefont {Marrazzo}, \citenamefont {Prandini}, \citenamefont
  {Bonf{\`{a}}}, \citenamefont {Atambo}, \citenamefont {Affinito},
  \citenamefont {Palummo}, \citenamefont {Molina-S{\'{a}}nchez}, \citenamefont
  {Hogan}, \citenamefont {Gr{\"{u}}ning}, \citenamefont {Varsano},\ and\
  \citenamefont {Marini}}]{Sangalli2019}%
  \BibitemOpen
  \bibfield  {author} {\bibinfo {author} {\bibfnamefont {D.}~\bibnamefont
  {Sangalli}}, \bibinfo {author} {\bibfnamefont {A.}~\bibnamefont {Ferretti}},
  \bibinfo {author} {\bibfnamefont {H.}~\bibnamefont {Miranda}}, \bibinfo
  {author} {\bibfnamefont {C.}~\bibnamefont {Attaccalite}}, \bibinfo {author}
  {\bibfnamefont {I.}~\bibnamefont {Marri}}, \bibinfo {author} {\bibfnamefont
  {E.}~\bibnamefont {Cannuccia}}, \bibinfo {author} {\bibfnamefont
  {P.}~\bibnamefont {Melo}}, \bibinfo {author} {\bibfnamefont {M.}~\bibnamefont
  {Marsili}}, \bibinfo {author} {\bibfnamefont {F.}~\bibnamefont {Paleari}},
  \bibinfo {author} {\bibfnamefont {A.}~\bibnamefont {Marrazzo}}, \bibinfo
  {author} {\bibfnamefont {G.}~\bibnamefont {Prandini}}, \bibinfo {author}
  {\bibfnamefont {P.}~\bibnamefont {Bonf{\`{a}}}}, \bibinfo {author}
  {\bibfnamefont {M.~O.}\ \bibnamefont {Atambo}}, \bibinfo {author}
  {\bibfnamefont {F.}~\bibnamefont {Affinito}}, \bibinfo {author}
  {\bibfnamefont {M.}~\bibnamefont {Palummo}}, \bibinfo {author} {\bibfnamefont
  {A.}~\bibnamefont {Molina-S{\'{a}}nchez}}, \bibinfo {author} {\bibfnamefont
  {C.}~\bibnamefont {Hogan}}, \bibinfo {author} {\bibfnamefont
  {M.}~\bibnamefont {Gr{\"{u}}ning}}, \bibinfo {author} {\bibfnamefont
  {D.}~\bibnamefont {Varsano}}, \ and\ \bibinfo {author} {\bibfnamefont
  {A.}~\bibnamefont {Marini}},\ }\href {\doibase 10.1088/1361-648X/ab15d0}
  {\bibfield  {journal} {\bibinfo  {journal} {J. Phys. Condens. Matter}\
  }\textbf {\bibinfo {volume} {31}},\ \bibinfo {pages} {325902} (\bibinfo
  {year} {2019})},\ \Eprint {http://arxiv.org/abs/1902.03837}
  {arXiv:1902.03837} \BibitemShut {NoStop}%
\bibitem [{\citenamefont {Madelung}\ \emph {et~al.}(1998)\citenamefont
  {Madelung}, \citenamefont {R{\"{o}}ssler},\ and\ \citenamefont
  {Schulz}}]{MadelungHandbook}%
  \BibitemOpen
  \bibinfo {editor} {\bibfnamefont {O.}~\bibnamefont {Madelung}}, \bibinfo
  {editor} {\bibfnamefont {U.}~\bibnamefont {R{\"{o}}ssler}}, \ and\ \bibinfo
  {editor} {\bibfnamefont {M.}~\bibnamefont {Schulz}},\ eds.,\ \href {\doibase
  10.1007/10681727_711} {\emph {\bibinfo {title}
  {{Landolt-B\"ornstein - Group III Condensed Matter
  \textperiodcentered Volume 41C: ``Non-Tetrahedrally Bonded
  Elements and Binary Compounds I'' in SpringerMaterials}}}}\ (\bibinfo
  {publisher} {Springer-Verlag Berlin Heidelberg},\ \bibinfo {address} {Berlin,
  Heidelberg},\ \bibinfo {year} {1998})\BibitemShut {NoStop}%
\bibitem [{\citenamefont {Wiedemeier}\ and\ \citenamefont
  {Siemers}(1977)}]{Wiedemeier1977}%
  \BibitemOpen
  \bibfield  {author} {\bibinfo {author} {\bibfnamefont {H.}~\bibnamefont
  {Wiedemeier}}\ and\ \bibinfo {author} {\bibfnamefont {P.~A.}\ \bibnamefont
  {Siemers}},\ }\href {\doibase 10.1002/zaac.19774310134} {\bibfield  {journal}
  {\bibinfo  {journal} {J. Inorg. Gen. Chem.}\ }\textbf {\bibinfo {volume}
  {431}},\ \bibinfo {pages} {299} (\bibinfo {year} {1977})}\BibitemShut
  {NoStop}%
\bibitem [{\citenamefont {Chattopadhyay}\ \emph {et~al.}(1987)\citenamefont
  {Chattopadhyay}, \citenamefont {Boucherle},\ and\ \citenamefont
  {Vonschnering}}]{Chattopadhyay1987}%
  \BibitemOpen
  \bibfield  {author} {\bibinfo {author} {\bibfnamefont {T.}~\bibnamefont
  {Chattopadhyay}}, \bibinfo {author} {\bibfnamefont {J.}~\bibnamefont
  {Boucherle}}, \ and\ \bibinfo {author} {\bibfnamefont {H.}~\bibnamefont
  {Vonschnering}},\ }\href {\doibase 10.1002/pssa.2211360106} {\bibfield
  {journal} {\bibinfo  {journal} {J. Phys. C Solid State Phys.}\ }\textbf
  {\bibinfo {volume} {20}},\ \bibinfo {pages} {1431} (\bibinfo {year}
  {1987})}\BibitemShut {NoStop}%
\bibitem [{\citenamefont {Brebrick}(1971)}]{Brebrick1971}%
  \BibitemOpen
  \bibfield  {author} {\bibinfo {author} {\bibfnamefont {R.~F.}\ \bibnamefont
  {Brebrick}},\ }\href@noop {} {\bibfield  {journal} {\bibinfo  {journal} {J.
  Phys. Chem. Solids}\ ,\ \bibinfo {pages} {551}} (\bibinfo {year}
  {1971})}\BibitemShut {NoStop}%
\bibitem [{\citenamefont {Shu}\ \emph {et~al.}(1987)\citenamefont {Shu},
  \citenamefont {Jaulmes}, \citenamefont {Ollitrault-Fichet},\ and\
  \citenamefont {Flahaut}}]{Shu1987}%
  \BibitemOpen
  \bibfield  {author} {\bibinfo {author} {\bibfnamefont {H.~W.}\ \bibnamefont
  {Shu}}, \bibinfo {author} {\bibfnamefont {S.}~\bibnamefont {Jaulmes}},
  \bibinfo {author} {\bibfnamefont {R.}~\bibnamefont {Ollitrault-Fichet}}, \
  and\ \bibinfo {author} {\bibfnamefont {J.}~\bibnamefont {Flahaut}},\
  }\href@noop {} {\bibfield  {journal} {\bibinfo  {journal} {J. Solid State
  Chem.}\ }\textbf {\bibinfo {volume} {69}},\ \bibinfo {pages} {48} (\bibinfo
  {year} {1987})}\BibitemShut {NoStop}%
\bibitem [{\citenamefont {Xu}\ \emph {et~al.}(2017)\citenamefont {Xu},
  \citenamefont {Xu},\ and\ \citenamefont {Zhu}}]{Xu2017}%
  \BibitemOpen
  \bibfield  {author} {\bibinfo {author} {\bibfnamefont {N.}~\bibnamefont
  {Xu}}, \bibinfo {author} {\bibfnamefont {Y.}~\bibnamefont {Xu}}, \ and\
  \bibinfo {author} {\bibfnamefont {J.}~\bibnamefont {Zhu}},\ }\href {\doibase
  10.1038/s41535-017-0054-3} {\bibfield  {journal} {\bibinfo  {journal} {npj
  Quantum Mater.}\ }\textbf {\bibinfo {volume} {2}},\ \bibinfo {pages} {51}
  (\bibinfo {year} {2017})}\BibitemShut {NoStop}%
\bibitem [{\citenamefont {Dr{\"{u}}ppel}\ \emph {et~al.}(2014)\citenamefont
  {Dr{\"{u}}ppel}, \citenamefont {Kr{\"{u}}ger},\ and\ \citenamefont
  {Rohlfing}}]{Druppel2014}%
  \BibitemOpen
  \bibfield  {author} {\bibinfo {author} {\bibfnamefont {M.}~\bibnamefont
  {Dr{\"{u}}ppel}}, \bibinfo {author} {\bibfnamefont {P.}~\bibnamefont
  {Kr{\"{u}}ger}}, \ and\ \bibinfo {author} {\bibfnamefont {M.}~\bibnamefont
  {Rohlfing}},\ }\href {\doibase 10.1103/PhysRevB.90.155312} {\bibfield
  {journal} {\bibinfo  {journal} {Phys. Rev. B}\ }\textbf {\bibinfo {volume}
  {90}},\ \bibinfo {pages} {155312} (\bibinfo {year} {2014})}\BibitemShut
  {NoStop}%
\bibitem [{\citenamefont {Lee}\ and\ \citenamefont {Yang}(2016)}]{Lee2016}%
  \BibitemOpen
  \bibfield  {author} {\bibinfo {author} {\bibfnamefont {C.~H.}\ \bibnamefont
  {Lee}}\ and\ \bibinfo {author} {\bibfnamefont {C.~K.}\ \bibnamefont {Yang}},\
  }\href {\doibase 10.1088/1367-2630/aa5132} {\bibfield  {journal} {\bibinfo
  {journal} {New J. Phys.}\ }\textbf {\bibinfo {volume} {18}},\ \bibinfo
  {pages} {123022} (\bibinfo {year} {2016})}\BibitemShut {NoStop}%
\bibitem [{\citenamefont {Tung}\ and\ \citenamefont {Cohen}(1969)}]{Tung1969}%
  \BibitemOpen
  \bibfield  {author} {\bibinfo {author} {\bibfnamefont {Y.~W.}\ \bibnamefont
  {Tung}}\ and\ \bibinfo {author} {\bibfnamefont {M.~L.}\ \bibnamefont
  {Cohen}},\ }\href {\doibase 10.1103/PhysRev.180.823} {\bibfield  {journal}
  {\bibinfo  {journal} {Phys. Rev.}\ }\textbf {\bibinfo {volume} {180}},\
  \bibinfo {pages} {823} (\bibinfo {year} {1969})}\BibitemShut {NoStop}%
\bibitem [{\citenamefont {Dalven}(1969)}]{Dalven1969}%
  \BibitemOpen
  \bibfield  {author} {\bibinfo {author} {\bibfnamefont {R.}~\bibnamefont
  {Dalven}},\ }\href {\doibase 10.1016/0020-0891(69)90022-0} {\bibfield
  {journal} {\bibinfo  {journal} {Infrared Phys.}\ }\textbf {\bibinfo {volume}
  {9}},\ \bibinfo {pages} {141} (\bibinfo {year} {1969})}\BibitemShut {NoStop}%
\bibitem [{\citenamefont {Allen}\ and\ \citenamefont
  {Heine}(1976)}]{Allen1976}%
  \BibitemOpen
  \bibfield  {author} {\bibinfo {author} {\bibfnamefont {P.~B.}\ \bibnamefont
  {Allen}}\ and\ \bibinfo {author} {\bibfnamefont {V.}~\bibnamefont {Heine}},\
  }\href {\doibase 10.1088/0022-3719/9/12/013} {\bibfield  {journal} {\bibinfo
  {journal} {J. Phys. C Solid State Phys.}\ }\textbf {\bibinfo {volume} {9}},\
  \bibinfo {pages} {2305} (\bibinfo {year} {1976})}\BibitemShut {NoStop}%
\bibitem [{\citenamefont {Allen}\ and\ \citenamefont
  {Cardona}(1981)}]{Allen1981}%
  \BibitemOpen
  \bibfield  {author} {\bibinfo {author} {\bibfnamefont {P.~B.}\ \bibnamefont
  {Allen}}\ and\ \bibinfo {author} {\bibfnamefont {M.}~\bibnamefont
  {Cardona}},\ }\href {\doibase 10.1103/PhysRevB.23.1495} {\bibfield  {journal}
  {\bibinfo  {journal} {Phys. Rev. B}\ }\textbf {\bibinfo {volume} {23}},\
  \bibinfo {pages} {1495} (\bibinfo {year} {1981})}\BibitemShut {NoStop}%
\bibitem [{\citenamefont {Cardona}\ and\ \citenamefont
  {Allen}(1983)}]{Cardona1983}%
  \BibitemOpen
  \bibfield  {author} {\bibinfo {author} {\bibfnamefont {M.}~\bibnamefont
  {Cardona}}\ and\ \bibinfo {author} {\bibfnamefont {P.}~\bibnamefont
  {Allen}},\ }\href@noop {} {\bibfield  {journal} {\bibinfo  {journal} {Phys.
  Rev. B}\ }\textbf {\bibinfo {volume} {27}},\ \bibinfo {pages} {4760}
  (\bibinfo {year} {1983})}\BibitemShut {NoStop}%
\bibitem [{\citenamefont {Murnaghan}(1944)}]{Murnaghan1944}%
  \BibitemOpen
  \bibfield  {author} {\bibinfo {author} {\bibfnamefont {F.~D.}\ \bibnamefont
  {Murnaghan}},\ }\href {\doibase 10.1073/pnas.30.9.244} {\bibfield  {journal}
  {\bibinfo  {journal} {Proc. Natl. Acad. Sci. USA}\ }\textbf {\bibinfo
  {volume} {30}},\ \bibinfo {pages} {244} (\bibinfo {year} {1944})}\BibitemShut
  {NoStop}%
\bibitem [{\citenamefont {Tyuterev}\ and\ \citenamefont
  {Vast}(2006)}]{Tyuterev2006}%
  \BibitemOpen
  \bibfield  {author} {\bibinfo {author} {\bibfnamefont {V.~G.}\ \bibnamefont
  {Tyuterev}}\ and\ \bibinfo {author} {\bibfnamefont {N.}~\bibnamefont
  {Vast}},\ }\href {\doibase 10.1016/j.commatsci.2005.08.012} {\bibfield
  {journal} {\bibinfo  {journal} {Comput. Mater. Sci.}\ }\textbf {\bibinfo
  {volume} {38}},\ \bibinfo {pages} {350} (\bibinfo {year} {2006})}\BibitemShut
  {NoStop}%
\bibitem [{\citenamefont {Dornhaus}\ \emph {et~al.}(1983)\citenamefont
  {Dornhaus}, \citenamefont {Nimtz},\ and\ \citenamefont
  {Schlicht}}]{Nimtz1983}%
  \BibitemOpen
  \bibfield  {author} {\bibinfo {author} {\bibfnamefont {R.}~\bibnamefont
  {Dornhaus}}, \bibinfo {author} {\bibfnamefont {G.}~\bibnamefont {Nimtz}}, \
  and\ \bibinfo {author} {\bibfnamefont {B.}~\bibnamefont {Schlicht}},\ }\href
  {\doibase 10.1007/bfb0044920} {\emph {\bibinfo {title} {{Narrow-gap
  semiconductors}}}}\ (\bibinfo  {publisher} {Springer, Berlin, Heidelberg},\
  \bibinfo {year} {1983})\BibitemShut {NoStop}%
\bibitem [{\citenamefont {Zasavitskii}\ \emph {et~al.}(2004)\citenamefont
  {Zasavitskii}, \citenamefont {{De Andrada E Silva}}, \citenamefont
  {Abramof},\ and\ \citenamefont {McCann}}]{Zasavitskii2004}%
  \BibitemOpen
  \bibfield  {author} {\bibinfo {author} {\bibfnamefont {I.~I.}\ \bibnamefont
  {Zasavitskii}}, \bibinfo {author} {\bibfnamefont {E.~A.}\ \bibnamefont {{De
  Andrada E Silva}}}, \bibinfo {author} {\bibfnamefont {E.}~\bibnamefont
  {Abramof}}, \ and\ \bibinfo {author} {\bibfnamefont {P.~J.}\ \bibnamefont
  {McCann}},\ }\href {\doibase 10.1103/PhysRevB.70.115302} {\bibfield
  {journal} {\bibinfo  {journal} {Phys. Rev. B}\ }\textbf {\bibinfo {volume}
  {70}},\ \bibinfo {pages} {115302} (\bibinfo {year} {2004})}\BibitemShut
  {NoStop}%
\bibitem [{\citenamefont {Li}\ \emph {et~al.}(2013)\citenamefont {Li},
  \citenamefont {Lin}, \citenamefont {Li}, \citenamefont {Li},\ and\
  \citenamefont {Liu}}]{Li2013e}%
  \BibitemOpen
  \bibfield  {author} {\bibinfo {author} {\bibfnamefont {Y.}~\bibnamefont
  {Li}}, \bibinfo {author} {\bibfnamefont {C.}~\bibnamefont {Lin}}, \bibinfo
  {author} {\bibfnamefont {H.}~\bibnamefont {Li}}, \bibinfo {author}
  {\bibfnamefont {X.}~\bibnamefont {Li}}, \ and\ \bibinfo {author}
  {\bibfnamefont {J.}~\bibnamefont {Liu}},\ }\href {\doibase
  10.1080/08957959.2013.848278} {\bibfield  {journal} {\bibinfo  {journal}
  {High Press. Res.}\ }\textbf {\bibinfo {volume} {33}},\ \bibinfo {pages}
  {713} (\bibinfo {year} {2013})}\BibitemShut {NoStop}%
\bibitem [{\citenamefont {Rousse}\ \emph {et~al.}(2005)\citenamefont {Rousse},
  \citenamefont {Klotz}, \citenamefont {Saitta}, \citenamefont
  {Rodriguez-Carvajal}, \citenamefont {McMahon}, \citenamefont {Couzinet},\
  and\ \citenamefont {Mezouar}}]{Rousse2005}%
  \BibitemOpen
  \bibfield  {author} {\bibinfo {author} {\bibfnamefont {G.}~\bibnamefont
  {Rousse}}, \bibinfo {author} {\bibfnamefont {S.}~\bibnamefont {Klotz}},
  \bibinfo {author} {\bibfnamefont {A.~M.}\ \bibnamefont {Saitta}}, \bibinfo
  {author} {\bibfnamefont {J.}~\bibnamefont {Rodriguez-Carvajal}}, \bibinfo
  {author} {\bibfnamefont {M.~I.}\ \bibnamefont {McMahon}}, \bibinfo {author}
  {\bibfnamefont {B.}~\bibnamefont {Couzinet}}, \ and\ \bibinfo {author}
  {\bibfnamefont {M.}~\bibnamefont {Mezouar}},\ }\href {\doibase
  10.1103/PhysRevB.71.224116} {\bibfield  {journal} {\bibinfo  {journal} {Phys.
  Rev. B}\ }\textbf {\bibinfo {volume} {71}},\ \bibinfo {pages} {224116}
  (\bibinfo {year} {2005})}\BibitemShut {NoStop}%
\bibitem [{\citenamefont {Pawbake}\ \emph {et~al.}(2019)\citenamefont
  {Pawbake}, \citenamefont {Bellin}, \citenamefont {Paulatto}, \citenamefont
  {B{\'{e}}neut}, \citenamefont {Biscaras}, \citenamefont {Narayana},
  \citenamefont {Late},\ and\ \citenamefont {Shukla}}]{Pawbake2019}%
  \BibitemOpen
  \bibfield  {author} {\bibinfo {author} {\bibfnamefont {A.}~\bibnamefont
  {Pawbake}}, \bibinfo {author} {\bibfnamefont {C.}~\bibnamefont {Bellin}},
  \bibinfo {author} {\bibfnamefont {L.}~\bibnamefont {Paulatto}}, \bibinfo
  {author} {\bibfnamefont {K.}~\bibnamefont {B{\'{e}}neut}}, \bibinfo {author}
  {\bibfnamefont {J.}~\bibnamefont {Biscaras}}, \bibinfo {author}
  {\bibfnamefont {C.}~\bibnamefont {Narayana}}, \bibinfo {author}
  {\bibfnamefont {D.~J.}\ \bibnamefont {Late}}, \ and\ \bibinfo {author}
  {\bibfnamefont {A.}~\bibnamefont {Shukla}},\ }\href {\doibase
  10.1103/PhysRevLett.122.145701} {\bibfield  {journal} {\bibinfo  {journal}
  {Phys. Rev. Lett.}\ }\textbf {\bibinfo {volume} {122}},\ \bibinfo {pages}
  {145701} (\bibinfo {year} {2019})}\BibitemShut {NoStop}%
\end{thebibliography}
%

\end{document}